\documentclass[11pt]{article}
\usepackage[utf8]{inputenc}
\usepackage{amsfonts}
\usepackage{amsmath}
\usepackage{amssymb}
\usepackage{indentfirst}
\usepackage{graphicx}
\usepackage[colorlinks]{hyperref}
\usepackage{cite}
\usepackage{colortbl}
\usepackage{bm}
\usepackage{microtype}
\usepackage[normalem]{ulem}

\numberwithin{equation}{section}
\allowdisplaybreaks

\begin{document}

\baselineskip=18pt 
\baselineskip 0.6cm
\begin{titlepage}
\vskip 4cm

\begin{center}
\textbf{\LARGE Three-dimensional non-relativistic Hietarinta supergravity }
\par\end{center}{\LARGE \par}

\begin{center}
	\vspace{1cm}
	\textbf{Patrick Concha}$^{\ast, \bullet}$,
	\textbf{Evelyn Rodríguez}$^{\ast, \bullet}$,
   \textbf{Sebastián Salgado}$^{\star}$
	\small
	\\[6mm]
	$^{\ast}$\textit{Departamento de Matemática y Física Aplicadas, }\\
	\textit{ Universidad Católica de la Santísima Concepción, }\\
\textit{ Alonso de Ribera 2850, Concepción, Chile.}
	\\[3mm]
 $^{\bullet}$\textit{Grupo de Investigación en Física Teórica, GIFT, }\\
	\textit{ Universidad Católica de la Santísima Concepción, }\\
\textit{ Alonso de Ribera 2850, Concepción, Chile.}
\\[3mm]
	$^{\star}$\textit{Instituto de Alta Investigación, Universidad de Tarapac\'{a},\\ Casilla 7D, Arica, Chile} \\[5mm]
	\footnotesize
	\texttt{patrick.concha@ucsc.cl},
	\texttt{erodriguez@ucsc.cl},
 	\texttt{sebasalg@gmail.com}
 
	\par\end{center}
\vskip 20pt
\centerline{{\bf Abstract}}
\medskip
\noindent In this work we present the non-relativistic regime of the Hietarinta gravity theory and its extension to supergravity. At the bosonic level, we derive the non-relativistic version of the Hietarinta model by employing a contraction process and addressing the non-degeneracy of the invariant metric. To incorporate supersymmetry, we apply the Lie algebra expansion method to obtain the non-relativistic formulation of $\mathcal{N}=2$ Hietarinta supergravity. Our results reveal that the non-relativistic Hietarinta theory encompasses the extended Bargmann (super)gravity as a special case, yet it differs significantly from other existing non-relativistic (super)gravity models. Furthermore, we generalize our analysis to include a cosmological constant term in the non-relativistic Hietarinta (super)gravity action and examine its effects on the torsion structure.

\end{titlepage}\newpage {\baselineskip=12pt \tableofcontents{}}

\section{Introduction}\label{intro}
The Hietarinta-type symmetries \cite{Hietarinta:1975fu} were introduced to generalize the D-dimensional Poincaré superalgebra by incorporating higher-spin generators. In its most fundamental formulation, the Hietarinta algebra in three spacetime dimensions serves as a natural extension of the Poincaré algebra, encompassing three spin-2 generators: the translational $P_A$, the Lorentz rotation $J_{A}$ and an additional generator $Z_A$ unique to the Hietarinta algebra \cite{Bansal:2018qyz,Chernyavsky:2019hyp,Chernyavsky:2020fqs,Concha:2023nou}.  Although an interchanging of the $P_A$ and $Z_A$ generators yields the Maxwell algebra \cite{Bacry:1970ye,Bacry:1970du,Schrader:1972zd,Gomis:2017cmt}, the Hietarinta basis provides a genuine extension of the Poincaré algebra, resulting in different physical interpretations. Notably, adding a symmetry-breaking term to the Hietarinta Chern-Simons (CS) action \cite{Chernyavsky:2020fqs} can reproduce the Minimal Massive Gravity theory \cite{Bergshoeff:2014pca}. Conversely, the Maxwell symmetry has been introduced to describe a Minkowski spacetime in presence of a constant electromagnetic field background \cite{Bacry:1970ye,Bacry:1970du,Schrader:1972zd,Gomis:2017cmt} and, along its generalizations, have proven useful in extending General Relativity in arbitrary spacetime dimensions through CS and Born-Infeld gravity models \cite{Edelstein:2006se,Izaurieta:2009hz,Concha:2014vka}. In three-dimensional spacetime, the Maxwell gravity CS theory is characterized by a Riemannian structure with vanishing torsion, and its implications have been extensively studied in \cite{Salgado:2014jka,Hoseinzadeh:2014bla,Aviles:2018jzw,Concha:2018zeb,Adami:2020xkm}.

In the context of supersymmetry, the Hietarinta CS supergravity model in three-dimensional spacetime was recently introduced in \cite{Concha:2023nou}. Unlike its bosonic counterpart, the supersymmetric extension of the Hietarinta algebra is not isomorphic to the Maxwell superalgebra \cite{Bonanos:2009wy,deAzcarraga:2012zv,deAzcarraga:2014jpa,Concha:2014xfa,Concha:2014tca} and exhibits a distinctly different structure. For comparison, the three-dimensional Maxwell CS supergravity \cite{Concha:2018jxx,Concha:2019icz} is a torsionless theory characterized by two Majorana supercharges, similar to the D'Auria-Fré superalgebra \cite{DAuria:1982uck} and the one proposed by Green \cite{Green:1989nn}. In contrast, the Hietarinta CS supergravity can be seen as an extension of the minimal Poincaré CS supergravity \cite{Achucarro:1987vz}. In this model, the Hietarinta gauge field not only introduces a non-vanishing torsion but alters the asymptotic structure of the theory. As demonstrated in \cite{Concha:2023nou}, after imposing appropriate boundary conditions, the asymptotic symmetry algebra of the Hietarinta CS supergravity theory is an extension of the $\mathfrak{bms}_3$ superalgebra \cite{Barnich:2014cwa,Barnich:2015sca,Caroca:2018obf}, featuring three central charges. This infinite-dimensional superalgebra represents a supersymmetric extension of the $\left(l=1\right)$-extended conformal Galilean Lie algebra\footnote{Supersymmetric extension of the $l$-extended conformal Galilean algebra can be found in \cite{Galajinsky:2017oiv,Galajinsky:2017rls,Galajinsky:2021hlr}.} \cite{Chernyavsky:2019hyp} and differs from the asymptotic symmetry of the minimal Maxwell supergravity \cite{Caroca:2019dds,Matulich:2023xpw}. 

Non-relativistic algebras have garnered increasing interest due to their relevance in condensed matter systems and non-relativistic effective field theories \cite{Son:2008ye,Balasubramanian:2008dm,Kachru:2008yh,Taylor:2008tg,Duval:2008jg,Bagchi:2009my,Hartnoll:2009sz,Bagchi:2009pe,Hoyos:2011ez,Son:2013rqa,Christensen:2013lma,Christensen:2013rfa,Abanov:2014ula,Hartong:2014oma,Hartong:2014pma,Hartong:2015wxa,Geracie:2015dea,Gromov:2015fda,Hartong:2015zia,Taylor:2015glc,Zaanen:2015oix,Devecioglu:2018apj}. At the gravity level, the non-relativistic regime of General Relativity was introduced by Cartan in \cite{Cartan1,Cartan2} and referred as Newton-Cartan (NC) gravity. Alternatively, General Relativity without cosmological constant can be expressed as a CS action in three spacetime dimensions based on the $\mathfrak{iso}\left(2,1\right)$ lie algebra \cite{Witten:1988hc} However, the non-relativistic limit of the $\mathfrak{iso}\left(2,1\right)$ CS gravity action suffers from degeneracy. Various strategies have been employed to address the degeneracy issue in the construction of non-relativistic gravity models. One approach involves to introduce central extensions of the non-relativistic algebra to resolve degeneracy problems. For instance, adding two central charges to the Galilean algebra, which reproduces the so-called extended Bargmann algebra \cite{Papageorgiou:2009zc,Bergshoeff:2016lwr}, allows for a non-degenerate invariant bilinear trace. Alternatively, applying a $U\left(1\right)$-enlargement of the relativistic algebra prior to taking the non-relativistic limit can reproduce a non-degenerate non-relativistic counterpart. Specifically, the extended Bargmann algebra \cite{Papageorgiou:2009zc,Bergshoeff:2016lwr,Matulich:2019cdo} can be derived from the $\mathfrak{iso}\left(2,1\right)\oplus\mathfrak{u}\left(1\right)^{2}$ algebra through a contraction process. Another method to avoid degeneracy in the non-relativistic regime is possible through the semigroup expansion procedure \cite{Izaurieta:2006zz}. As it was shown in \cite{Gomis:2019nih,Concha:2023bly}, a non-relativistic expansion can be applied using an appropriate semigroup without requiring a $U\left(1\right)$-enlargement of the relativistic algebra. Consequently, the extended Bargmann algebra can also be seen as an expansion of the Poincaré algebra via a suitable semigroup \cite{Gomis:2019nih,Concha:2023bly}.

On the other hand, non-relativistic supergravity theories have only been approached recently mainly due to the challenges in applying a proper non-relativistic limit in presence of supersymmetry. Despite these difficulties, there have been various successful attempts in the literature to construct non-relativistic supergravity actions \cite{Andringa:2013mma, Bergshoeff:2015ija, Bergshoeff:2016lwr,Ozdemir:2019orp, deAzcarraga:2019mdn, Ozdemir:2019tby, Concha:2019mxx, Concha:2020tqx, Concha:2020eam,Concha:2021jos,Concha:2021llq,Ravera:2022buz,Bergshoeff:2022iyb,Concha:2023qfj}. Of particular interest is the extension of the non-relativistic expansion \cite{Gomis:2019nih,Concha:2023bly} to supersymmetric theories. This approach not only facilitates the construction of well-defined three-dimensional non-relativistic supergravity models but also prevents the degeneracy problem \cite{Concha:2020tqx,Concha:2021jos,Concha:2021llq}. In this context, the extended Bargmann supergravity \cite{Bergshoeff:2016lwr} along novel generalizations have been derived by expanding $\mathcal{N}=2$ relativistic superalgebras. The use of $\mathcal{N}=2$ ensures that the expanded non-relativistic superalgebra yields a genuine supergravity action, in which the time translational generators can be expressed as bilinear combinations of supersymmetry generators \cite{Bergshoeff:2022iyb}.

To our knowledge, the exploration of a non-relativistic regime of the Hietarinta (super)gravity theory remains unknown. While a non-relativistic version of the Hietarinta CS action can be derived from the Maxwellian extended Bargmann action \cite{Aviles:2018jzw} by interchanging the corresponding gauge field \cite{Penafiel:2019czp}, this process cannot be naively applied in the presence of supersymmetry due to differences between their relativistic counterparts. In this work, we introduce a new non-relativistic (super)gravity theory that generalizes the extended Bargmann model without introducing a cosmological constant. The resulting CS (super)gravity action is based on a non-relativistic version of the Hietarinta (super)algebra. At the bosonic level, we apply the standard non-relativistic limit to a $U\left(1\right)$-enlargement of the Hietarinta algebra, while its supersymmetric extension requires using the expansion method outlined in \cite{Concha:2020tqx,Concha:2021jos,Concha:2021llq}. Interestingly, the resulting (super)algebra can also be obtained by contracting known non-relativistic (super)algebras. Furthermore, we demonstrate that the Hietarinta extended Bargmann (super)gravity model can alternatively be derived as the vanishing cosmological constant limit of a more general non-relativistic (super)gravity theory.

The organization of the paper is as follows: In section \ref{sec2} we review the relativistic Hietarinta gravity and introduce its $U\left(1\right)$-enlargement. Sections \ref{sec3} and \ref{sec4} present our main results. Section \ref{sec3} discusses the non-relativistic limit of the Hietarinta algebra and introduces the Hietarinta extended Bargmann gravity theory. We analyze and address the degeneracy issue by considering the Hietarinta $\oplus\,\mathfrak{u}\left(1\right)^{3}$ algebra as its relativistic counterpart. Section \ref{sec4} is dedicated to the supersymmetric extension of the Hietarinta extended Bargmann gravity, employing the semigroup expansion method. We also explore the incorporation of a cosmological constant into the non-relativistic Hietarinta supergravity theory. Finally, Section \ref{sec5} concludes with comments on potential future developments.

\section{Relativistic Hietarinta gravity}\label{sec2}
\subsection{ Hietarinta CS gravity}
In this section we briefly review the three-dimensional gravity theory based on the so-called Hietarinta algebra \cite{Chernyavsky:2020fqs}, as well as, its $U\left(1\right)$-enlargement required for a proper non-relativistic limit. In its simplest form, the Hietarinta algebra in three spacetime dimensions contains three spin-2 generators which obey the following structure \cite{Chernyavsky:2020fqs}
\begin{align}
    \left[ \tilde{J}_{A},\tilde{J}_{B}\right] &=\epsilon _{ABC}\tilde{J}^{C}\,,& \left[ \tilde{J}_{A},\tilde{P}_{B}\right] &=\epsilon _{ABC}\tilde{P}^{C}\,, \notag \\
\left[ \tilde{J}_{A},\tilde{Z}_{B}\right] &=\epsilon _{ABC}\tilde{Z}^{C}\,,  & \left[ \tilde{Z}_{A},\tilde{Z}_{B}\right] &=\epsilon _{ABC}\tilde{P}^{C}\,, \label{Hieta}
\end{align} 
where $\tilde{J}_{A},\tilde{P}_{A}$ and $\tilde{Z}_{A}$ correspond to Lorentz rotations, translations and an additional vector generator, respectively. Here $A,B,\cdots=0,1,2$ denote Lorentz indices which are lowered and raised with the Minkowski metric $\eta_{AB}=\left(-1,1,1\right)$ and $\epsilon_{ABC}$ is the Levi-Civita symbol. This algebra has also been denoted as the Hietarinta/Maxwell algebra \cite{Chernyavsky:2020fqs} due to its isomorphism with the Maxwell algebra \cite{Bacry:1970ye,Bacry:1970du,Schrader:1972zd,Gomis:2017cmt}, in which the role of $\tilde{Z}_{A}$ and $\tilde{P}_{A}$ is interchanged. The Hietarinta algebra can be seen as an extension of the Poincaré algebra unlike the Maxwell symmetry, which corresponds to an extension and deformation of the Poincaré one.

The Hietarinta algebra \eqref{Hieta} admits a non-degenerate invariant tensor whose non-vanishing components are given by \cite{Chernyavsky:2020fqs}
\begin{align}
\left\langle \tilde{J}_{A}\tilde{J}_{B}\right\rangle &=\tilde{\alpha} _{0}\eta _{AB}\,,  &  \left\langle \tilde{J}_{A}\tilde{P}_{B}\right\rangle &=\tilde{\alpha}_{1}\eta _{AB}\,, & \left\langle \tilde{Z}_{A}\tilde{Z}_{B}\right\rangle &=\tilde{\alpha} _{1}\eta _{AB}\,,  \notag \\
\left\langle \tilde{J}_{A}\tilde{Z}_{B}\right\rangle &=\tilde{\alpha} _{2}\eta _{AB}\,,\label{ITHie}
\end{align}
where $\tilde{\alpha}_0$, $\tilde{\alpha}_1$ and $\tilde{\alpha}_2$ are arbitrary constants.
The gauge connection one-form $A$ reads
\begin{equation}
A=W ^{A}\tilde{J}_{A}+E^{A}\tilde{P}_{A}+K^{A}\tilde{Z}_{A}
\,,  \label{1fP}
\end{equation}%
where $W^{A}$ corresponds the spin connection one-form, $E^{A}$ is the
the dreibein one-form and $K^{A}$ refers to the gauge field one-form associated with the vector generator $\tilde{Z}_A$.

The CS action invariant under the Hietarinta algebra \eqref{Hieta} is obtained considering the gauge
connection one-form (\ref{1fP}) along the non-vanishing components of the invariant tensor (\ref{ITHie}) in the general expression of a three-dimensional CS action
\begin{equation}
I[A]=\frac{k}{4\pi }\int_{\mathcal{M}}\left\langle AdA+\frac{2}{3}
A^{3}\right\rangle \,,  \label{CSaction}
\end{equation}
where $k$ is the CS level of the theory related to the gravitational constant $G$ through $k=\frac{1}{4G}$. 

Then, the CS gravity action based on the Hietarinta algebra \eqref{Hieta} reads \cite{Chernyavsky:2020fqs}
\begin{eqnarray}
I_{\mathcal{H}} &=&\frac{k}{4\pi }\int \tilde{\alpha}_{0}\left( \,W
^{A}dW _{A}+\frac{1}{3}\,\epsilon _{ABC}W ^{A}W ^{B}W
^{C}\right)  \notag \\
&&+\tilde{\alpha}_{1}\left( 2E^{A}R_{A}+K^{A}\mathcal{F}_{A} \right) \,+2\tilde{\alpha}_{2} K ^{A} R_{A} \,,  \label{HCS}
\end{eqnarray}
where 
\begin{eqnarray}
R^{A} &=&dW ^{A}+\frac{1}{2}\epsilon ^{ABC}W _{B}W _{C}\,,
\notag \\
\mathcal{F}^{A} &=&dK^{A}+\epsilon ^{ABC}W _{B}K _{C}
\,, \label{bosc}
\end{eqnarray}
are the curvature two-forms for the spin-connection and the $K^{A}$ gauge field, respectively. This action can be seen as an extension of the three-dimensional Poincaré CS gravity action. We can see that the term proportional to $\tilde{\alpha}_{0}$ contains the gravitational "exotic" CS Lagrangian \cite{Witten:1988hc,Zanelli:2005sa}. On the other hand, the $K^{A}$ gauge field appears explicitly along the Einstein-Hilbert term in the $\tilde{\alpha}_{1}$ sector and along the $\tilde{\alpha}_{2}$ term. Let us note that the Maxwell CS gravity action \cite{Salgado:2014jka,Hoseinzadeh:2014bla,Aviles:2018jzw,Concha:2018zeb} is recovered when the role of the dreiben $E^{A}$ and the $K^{A}$ gauge field is interchanged. Although both Maxwell and Hietarinta algebras are related through a trivial interchanging of generators, they possess different physical implications. Indeed, one can check that the field equations coming from the Hietarinta CS gravity action are given by the vanishing of the curvature two-forms \eqref{bosc} and the vanishing of the curvature for the dreiben:
\begin{eqnarray}
\mathcal{T}^{A}&=&dE^{A}+\epsilon^{ABC}W_{B}E_{C}+\epsilon^{ABC}K_{B}K_{C}\,.
\end{eqnarray}
One can see that the Hietarinta equations of motion differ from the field equations for the Maxwellian case \cite{Salgado:2014jka,Hoseinzadeh:2014bla,Aviles:2018jzw,Concha:2018zeb}:
\begin{eqnarray}
{T}^{A}&=&dE^{A}+\epsilon^{ABC}W_{B}E_{C}=0\,,\notag \\
F^{A}&=&dK^{A}+\epsilon^{ABC}W_{B}K_{C}+\epsilon^{ABC}E_{B}E_{C}=0\,.
\end{eqnarray}
The presence of the $K^{A}$ gauge field in the Hietarinta gravity not only turns on the torsion $T^{A}$ but also, as it was shown in \cite{Concha:2023nou}, modifies the asymptotic symmetry. In particular, after imposing suitable boundary conditions, the asymptotic symmetry algebra for the three-dimensional Hietarinta gravity theory is given by an extension of the $\mathfrak{bms}_{3}$ algebra which can be seen as the extended l-conformal Galilean algebra for $l=1$ \cite{Chernyavsky:2019hyp}.

\subsection{$U(1)$ enlargement}
As we shall see in the next section, a straightforward non-relativistic limit of the action \eqref{HCS} leads to degeneracy. One way to circumvent the degeneracy problem occurring in the contraction process is to add $\mathfrak{u}\left(1\right)$ generators in the relativistic algebra. Then, inspired by the Maxwell case treated in \cite{Aviles:2018jzw}, we will include three extra $U(1)$ one-form gauge fields, $M, T$ and $S$ at the relativistic level, such that the new connection one-form is
\begin{equation}
A=W ^{A}\tilde{J}_{A}+E^{A}\tilde{P}_{A}+K^{A}\tilde{Z}_{A}+MY_1+SY_2+TY_3
\,. \label{1fPe}
\end{equation} 
The Hietarinta$\,\oplus\, \mathfrak{u}\left(1\right)^{3}$ algebra admits the invariant tensor \eqref{ITHie} along the following non-vanishing components
\begin{align}
\left\langle Y_{1}Y_{2}\right\rangle &=\left\langle Y_{3}Y_{3}\right\rangle=\tilde{\alpha}_{1}\,,  &  \left\langle Y_{2}Y_{3}\right\rangle &=\tilde{\alpha}_{2}\,,&  \left\langle Y_{2}Y_{2}\right\rangle &=\tilde{\alpha}_{0}\,.\label{ITHiee}
\end{align}
Therefore, the Hietarinta$\,\oplus \,\mathfrak{u}\left(1\right)^{3}$ action becomes
\begin{eqnarray}
I_{\mathcal{H}\oplus\mathfrak{u}(1)^{3}} &=&\frac{k}{4\pi }\int \tilde{\alpha}_{0}\left( \,W
^{A}dW _{A}+\frac{1}{3}\,\epsilon _{ABC}W ^{A}W ^{B}W
^{C}+SdS\right)  \notag \\
&&+\tilde{\alpha}_{1}\left( 2E^{A}R_{A}+K^{A}\mathcal{F}_{A} +2MdS+TdT\right) \,+2\tilde{\alpha}_{2}\left( K ^{A} R_{A} +TdS\right)\,,  \label{HCSu1}
\end{eqnarray}
where we have considered the gauge connection one-form \eqref{1fPe} along the invariant tensor \eqref{ITHie} and \eqref{ITHiee} in the general expression of the CS form \eqref{CSaction}.

\section{ Non-relativistic Hietarinta CS gravity}\label{sec3}
We now consider the IW contraction of the previously introduced Hietarianta algebra \eqref{Hieta}. To this end we decompose the spacetime index $A$ as follows
\begin{align}
    A&\rightarrow \left(0,a\right)\,, \label{dec}
\end{align}
with $a=1,2$. We first show that a straightforward non-relativistic limit prevents to construct a proper non-relativistic CS action without degeneracy. The inclusion of three $U\left(1\right)$ gauge fields is required in the contraction process to avoid degeneracy and obtain a well-defined non-relativistic action. The corresponding non-relativistic algebra can be seen as an extension of the extended Bargmann algebra \cite{Papageorgiou:2009zc,Bergshoeff:2016lwr}.

\subsection{Non-relativistic limit of the Hietarinta algebra}
A non-relativistic Hietarinta algebra can be obtained as a contraction of the relativistic Hietarinta algebra \eqref{Hieta}. Such non-relativistic contraction is performed by rescaling the Hietarinta generators with a $\xi$ parameter as follows:\footnote{The $\xi$ parameter is related to the speed of light $c$ as $\xi\rightarrow 1/c$.},
\begin{align}
    \tilde{J}_{0}&= J\,, & \tilde{P}_{0}&= H\,, & \tilde{Z}_{0}&= Z\,, \notag \\
    \tilde{J}_{a}&= \xi G_{a}\,, & \tilde{P}_{a}&= \xi P_{a}\,, &
    \tilde{Z}_{a}&= \xi Z_{a}\,,
\end{align}
and by applying the non-relativistic limit $\xi\rightarrow \infty$. The contracted algebra corresponds to the non-relativistic counterpart of the Hietarinta algebra. The generators in this non-relativistic framework satisfy the following non-vanishing commutation relations:
\begin{align}
\left[  J,G_{a}\right]   & =\epsilon_{ab}G_{b}\,, &
\left[  H,G_{a}\right]   &  =\epsilon_{ab}P_{b}\,, &
\left[  J,P_{a}\right]   &  =\epsilon_{ab}P_{b}\,,\notag \\
\left[  J,Z_{a}\right]   &  =\epsilon_{ab}Z_{b}\,, &
\left[  Z,G_{a}\right]   &  =\epsilon_{ab}Z_{b}\,, &
\left[  Z,Z_{a}\right]   &  =\epsilon_{ab}P_{b}\,, \label{HGal}
\end{align}
where $\epsilon_{ab}\equiv\epsilon_{0ab}$ and $\epsilon^{ab}\equiv\epsilon^{0ab}$. The non-relativistic algebra \eqref{HGal}, which we have denoted as the Hietarinta Galilei ($\mathfrak{hgal}$) algebra, corresponds to an extension of the Galilei algebra spanned by $\{J, H, G_a, P_a \}$. Let us note that the $\mathfrak{hgal}$ algebra can be alternatively obtained from three copies of the Poincaré algebra $\mathfrak{iso}\left(1,1\right)$,
\begin{align}
\left[  J^{\pm},P^{\pm}_{a}\right]   & =\epsilon_{ab}P^{\pm}_{b}\,, &
\left[  \hat{J},\hat{P}_{a}\right]   &  =\epsilon_{ab}\hat{P}_{b}\,, 
\end{align}
Indeed, the $\mathfrak{hgal}$ algebra is recovered by considering the following redefinition
\begin{align}
G_{a}&=\hat{P}_{a}+P^{+}_{a}+P^{-}_{a}\,, & P_{a}&=\frac{1}{\sigma^{2}}\left(P^{+}_{a}+P^{-}_{a}\right)\,, & Z_a&=\frac{1}{\sigma}\left(P^{+}_{a}-P^{-}_{a} \right)\,,\notag \\
J&=\hat{J}+J^{+}+J^{-}\,, & H&=\frac{1}{\sigma^2}\left( J^{+}+J^{-}\right)\,, & Z&=\frac{1}{\sigma}\left( J^{+}-J^{-}\right)\,,
\end{align}
along the limit $\sigma\rightarrow\infty$.
Nonetheless, as its Galilean subalgebra, the $\mathfrak{hgal}$ algebra does not admit a non-degenerate bilinear invariant form:
\begin{align}
    \langle J J \rangle &= -\mu_0\,, & \langle J H \rangle &= -\mu_1\,, & \langle J Z \rangle &= \langle H H \rangle = -\mu_2\,. \label{IThgal}
\end{align}
The non-degeneracy of the action is a suitable feature since it ensures a kinetic term for each gauge field and yields to the vanishing of the curvatures as field equations. One could construct a CS action for the $\mathfrak{hgal}$ algebra considering the non-vanishing components of the invariant tensor \eqref{IThgal} and the one-form gauge connection for the $\mathfrak{hgal}$ algebra,
\begin{equation}
    A=\omega J +\omega_a G_a + \tau H + e^{a}P_{a} + k Z + k^{a} Z_{a}\,,
\end{equation}
However, one can easily check from the invariant tensor \eqref{IThgal} that there is no contribution of the spatial gauge field $\{\omega^{a},e^{a},k^{z}\}$ in the action.

\subsection{Hietarinta extended Bargmann algebra and non-relativistic gravity action}
A non-relativistic Hietarinta algebra without degeneracy can be obtained by considering the IW contraction of the previously introduced Hietarinta $\oplus\, \mathfrak{u}\left(1\right)^{3}$ algebra. For this purpose, we perform the identification of the relativistic generators with the non-relativistic ones through the $\xi$ parameter as
\begin{align}
    J_{0}&=  \frac{J}{2}+\xi^2 S \,,&  J_{a}&=\xi G_{a} \,,  & Y_{2}&= \frac{J}{2}-\xi^2 S\,, \notag\\
   {P}_{0}&= \frac{H}{2\xi^2}+
   M\,, &P_{a}&= \frac{P_{a}}{\xi}\,, & Y_{1}&=\frac{H}{2\xi^2}- M\,,\notag \\
   {Z}_{0}&= \frac{Z}{2\xi}+\xi T\,, &Z_{a}&= Z_{a}\,, & Y_{3}&=\frac{Z}{2\xi}-\xi T\,. \label{NRlim}
\end{align}
Then, after applying the non-relativistic limit $\xi\rightarrow\infty$, the non-relativistic Hietarinta algebra reads
\begin{align}
\left[  J,G_{a}\right]   &  =\epsilon_{ab}G_{b}\,, &
\left[  J,P_{a}\right]   &  =\epsilon_{ab}P_{b}\,,\notag \\
\left[  G_{a},G_{b}\right]   &  =-\epsilon_{ab}S\,,&
\left[  H,G_{a}\right]   &  =\epsilon_{ab}P_{b}\,,\notag \\
\left[  G_{a},P_{b}\right]   &  =-\epsilon_{ab}M\,,&
\left[  J,Z_{a}\right]   &  =\epsilon_{ab}Z_{b}\,,\notag \\
\left[  G_{a},Z_{b}\right]   &  =-\epsilon_{ab}T\,, & \left[  Z,G_{a}\right]   &  =\epsilon_{ab}Z_{b}\,, \notag
\\
\left[  Z_{a},Z_{b}\right]   &  =-\epsilon_{ab}M\,,& 
\left[  Z,Z_{a}\right]   &  =\epsilon_{ab}P_{b}\,.\label{heb}
\end{align}
The obtained non-relativistic algebra, which we have denoted as the Hietarinta extended Bargmann ($\mathfrak{heb}$) algebra, has been first introduced in \cite{Penafiel:2019czp} by expanding the so-called Nappi-Witten algebra \cite{Nappi:1993ie,Figueroa-OFarrill:1999cmq}. One can note that the $\mathfrak{heb}$ algebra contains the extended Bargmann algebra \cite{Papageorgiou:2009zc,Bergshoeff:2016lwr} spanned by $\{J, G_a, H, P_a, S, M\}$ as a subalgebra and can be seen as a double central extension of the $\mathfrak{hgal}$ algebra \eqref{HGal} introduced previously. The presence of the central charges $S$ and $M$ would ensure the non-degeneracy of the invariant bilinear trace. As it was mentioned in \cite{Penafiel:2019czp}, this non-relativistic Hietarinta algebra can also been obtained from the Maxwellian Exotic Bargmann ($\mathfrak{meb}$) symmetry introduced in \cite{Aviles:2018jzw} by interchanging the generators as $P_a\leftrightarrow Z_a, H\leftrightarrow Z, M\leftrightarrow T$. 

An alternative procedure to obtain the $\mathfrak{heb}$ algebra is through the contraction of three copies of the Nappi-Witten algebra\footnote{The Nappi-Witten algebra can be seen as a central extension of the two-dimensional Poincaré algebra \cite{Nappi:1993ie,Figueroa-OFarrill:1999cmq}.},
\begin{align}
\left[  J^{\pm},P^{\pm}_{a}\right]   & =\epsilon_{ab}P^{\pm}_{b}\,, & \left[P^{\pm}_{a},P^{\pm}_{b}\right]&=-\epsilon_{ab}T^{\pm}\,, \notag\\
\left[  \hat{J},\hat{P}_{a}\right]   &  =\epsilon_{ab}\hat{P}_{b}\,, &\left[\hat{P}_{a},\hat{P}_{b} \right]&=-\epsilon_{ab} \hat{T}\,.
\end{align}
The $\mathfrak{heb}$ algebra \eqref{heb} is obtained by considering the redefinition of the Nappi-Witten generators as
\begin{align}
    G_{a}&=\hat{P}_{a}+P^{+}_{a}+P^{-}_{a}\,, & P_{a}&=\frac{1}{\sigma^{2}}\left(P^{+}_{a}+P^{-}_{a}\right)\,, & Z_a&=\frac{1}{\sigma}\left(P^{+}_{a}-P^{-}_{a} \right)\,,\notag \\
J&=\hat{J}+J^{+}+J^{-}\,, & H&=\frac{1}{\sigma^2}\left( J^{+}+J^{-}\right)\,, & Z&=\frac{1}{\sigma}\left( J^{+}-J^{-}\right)\,,\notag \\
S&=\hat{T}+T^{+}+T^{-}\,,& M&=\frac{1}{\sigma^{2}}\left(T^{+}+T^{-}\right)\,, & T&=\frac{1}{\sigma}\left(T^{+}-T^{-}\right)\,,
\end{align}
along the limit $\sigma\rightarrow\infty$. Alternatively, the $\mathfrak{heb}$ algebra can also be derived from the direct sum of the extended Newton-Hooke algebra \cite{Aldrovandi:1998im,Gibbons:2003rv,Brugues:2006yd,Alvarez:2007fw,Papageorgiou:2010ud,Duval:2011mi,Duval:2016tzi,Concha:2023bly}:
\begin{align}
\left[  \texttt{J},\texttt{G}_{a}\right]   &  =\epsilon_{ab}\texttt{G}_{b}\,, &
\left[  \texttt{J},\texttt{P}_{a}\right]   &  =\epsilon_{ab}\texttt{P}_{b}\,,\notag \\
\left[  \texttt{G}_{a},\texttt{G}_{b}\right]   &  =-\epsilon_{ab}\texttt{S}\,,&
\left[  \texttt{H},\texttt{G}_{a}\right]   &  =\epsilon_{ab}\texttt{P}_{b}\,,\notag \\
\left[  \texttt{G}_{a},\texttt{P}_{b}\right]   &  =-\epsilon_{ab}\texttt{M}\,,&
\left[  \texttt{H},\texttt{P}_{a}\right]   &  =\frac{1}{\ell^2}\epsilon_{ab}\texttt{G}_{b}\,,\notag \\
\left[  \texttt{P}_{a},\texttt{P}_{b}\right]   &  =-\frac{1}{\ell^{2}}\epsilon_{ab}\texttt{S}\,.\label{enh}
\end{align}
and the Nappi-Witten one \cite{Nappi:1993ie,Figueroa-OFarrill:1999cmq} spanned by $\{\Bar{J},\Bar{P}_{a},\Bar{T}\}$. In particular, the $\mathfrak{heb}$ algebra is recovered after considering the redefinition of the generators, 
\begin{align}
G_{a}&=\Bar{P}_{a}+\texttt{G}_{a}\,, & P_{a}&=\frac{1}{\ell^{2}}\texttt{G}_{a}\,, & Z_{a}&=\texttt{P}_{a}\,,\notag \\
J&=\Bar{J}+\texttt{J}\,, & H&=\frac{1}{\ell^2}\texttt{J}\,, & Z&=\texttt{H}\,,\notag \\
S&=\Bar{T}+\texttt{S}\,, & M&=\frac{1}{\ell^2}\texttt{S}\,, & T&=\texttt{M}\,,
\end{align}
and performing the limit $\ell\rightarrow\infty$.  Although diverse contractions can be applied to obtain the $\mathfrak{heb}$ algebra, we shall focus on the non-relativistic limit of the Hietarinta $\oplus\,\mathfrak{u}\left(1\right)^{3}$ algebra for constructing the corresponding CS action.

A non-relativistic CS gravity action based on the $\mathfrak{heb}$ symmetry \eqref{heb} can be constructed by considering the gauge connection one-form taking values on the Hietarinta extended Bargamnn algebra:
\begin{eqnarray}
    A&=&\omega J +\omega_a G_a + \tau H + e^{a}P_{a} + k Z + k^{a} Z_{a}+s S +m M+ t T\,. \label{Aheb}
\end{eqnarray}
Here $\omega$, $\omega_{a}$, $\tau$ and $e_{a}$ are the time and spatial components of the spin-connection and dreibein, respectively. On the other hand, $s$, $m$ and $t$ are the gauge fields associated to the central charges $S$, $M$ and $T$, respectively. Unlike the $\mathfrak{hgal}$ algebra \eqref{HGal} previously introduced, the $\mathfrak{heb}$ algebra admits a non-degenerate invariant bilinear trace whose non-vanishing components are given by
\begin{align}
    \langle J S \rangle&=-\alpha_0\,, & \langle G_{a}G_{b}\rangle &=\alpha_0 \delta_{ab}\,, & \langle J M \rangle &=-\alpha_1\,,\notag \\
    \langle H S \rangle &=-\alpha_1 \,, & \langle G_{a} P_{b} \rangle &=\alpha_1 \delta_{ab}\,, & \langle T Z \rangle &=-\alpha_1 \,,\notag \\
    \langle J T \rangle &=-\alpha_2 \,, & \langle Z_{a}Z_{b} \rangle&=\alpha_1 \delta_{ab}\,,  \notag \\  \langle Z S \rangle &=-\alpha_2\,, & \langle G_{a}Z_{b}\rangle &=\alpha_2\delta_{ab}\,. \label{ITHEB}
\end{align}
Here, $\alpha_0$, $\alpha_1$ and $\alpha_2$ are related to the relativistic Hietarianta$\oplus\,\mathfrak{u}\left(1\right)^{3}$ parameters as follows,
\begin{align}
    \tilde{\alpha}_0&=\xi^{2} \alpha_0\,, & \tilde{\alpha}_{1}&=\alpha_1\,, & \tilde{\alpha}_{2}&=\xi \alpha_2\,. \label{Ident}
\end{align}
The non-vanishing components of the invariant tensor for the $\mathfrak{heb}$ algebra are obtained from the relativistic ones \eqref{ITHie} considering the identification \eqref{NRlim} and \eqref{Ident} along the limit $\xi\rightarrow\infty$. Then, by replacing the gauge connection one-form \eqref{Aheb} and the invariant tensor \eqref{ITHEB} in the general expression CS expression \eqref{CSaction} we get
\begin{eqnarray}
    I_{\mathfrak{heb}}&=&\int \alpha_0\left[\omega_a R^{a}\left(\omega^{b}\right)-2 s R\left(\omega\right)\right]\notag\\ &&\ +\alpha_1\left[2e_aR^{a}\left(\omega^{b}\right)-2m R\left(\omega\right)-2\tau R\left(s\right)+k_{a}R^{a}\left(k^{b}\right)-2k R\left(t\right)\right]\notag\\
    &&\ +\alpha_2\left[2k_{a}R^{a}\left(\omega^{b}\right)-2t R\left(\omega\right)-2k R\left(s\right)\right]\,,\label{CSheb}
\end{eqnarray}
where
\begin{align}
    R\left(\omega\right)&=d\omega\,,\notag\\
R^{a}\left(\omega^{b}\right)&=d\omega^{a}+\epsilon^{ac}\omega\omega_c\,,\notag\\
R^{a}\left(k^{b}\right)&=dk^{a}+\epsilon^{ac}\omega k_{c}\,,\notag\\
R\left(s\right)&=ds+\frac{1}{2}\epsilon^{ac}\omega_{a}\omega_{c}\,, \notag\\
R\left(t\right)&=dt+\epsilon^{ac}\omega_{a} k_{c}\,. \label{acurv}
\end{align}
The CS action \eqref{CSheb} describes the Hietarinta extended Bargmann gravity theory \cite{Penafiel:2019czp} and reproduces the extended Bargmann one for $k=k^{a}=t=0$. One can notice that the exotic term along $\alpha_0$, which is the non-relativistic counterpart of the exotic gravitational CS term \cite{Witten:1988hc}, coincides with the exotic sector of the extended Bargmann gravity theory \cite{Concha:2023bly}. On the other hand, the term proportional to $\alpha_2$ corresponds to a new sector involving the extra Hietarinta gauge fields $k$ and $k_a$. 
The presence of these gauge fields modifies the extended Bargmann field equations to the vanishing of the following curvature two-forms:
\begin{align}
\texttt{R}\left(\omega\right)&=R\left(\omega\right)\,,&
\texttt{R}^{a}\left(\omega^{b}\right)&=R^{a}\left(\omega^{b}\right)\,,\notag\\
\texttt{R}\left(\tau\right)&=d\tau\,,&
\texttt{R}^{a}\left(e^{b}\right)&=de^{a}+\epsilon^{ac}\omega e_{c}+\epsilon^{ac}\tau\omega_{c}+\epsilon^{ac}k k_{c}\,,\notag\\
\texttt{R}\left(s\right)&=R\left(s\right)\,,&
\texttt{R}\left(m\right)&=dm+\epsilon^{ac}\omega_{a}e_{c}+\frac{1}{2}\epsilon^{ac}k_{a}k_{c}\,,\notag\\
\texttt{R}\left(k\right)&=dk\,,&
\texttt{R}^{a}\left(k^{b}\right)&=R^{a}\left(k^{b}\right)+\epsilon^{ac}k\omega_{c}\,,\notag\\
\texttt{R}\left(t\right)&=R\left(t\right)\,.\label{hebeom}
\end{align}
It would be interesting to study whether the previous action can be derived within the NC formalism. This construction, however, could require extending the standard NC geometry based on the usual Bargmann algebra. As pointed out in \cite{Hansen:2019pkl}, it was emphasized that the NC geometry, which allows to geometrize the Poisson equation of Newtonian gravity, requires modifications to construct an action principle for Newtonian gravity. The underlying symmetry in this case extends the Galilean generators $\{H,P_{a},G_{a},J_{ab}$\} by introducing the $\{N,T_{a},B_{a},S_{ab}\}$ generators, which satisfy the so-called Newtonian or Post-Newtonian algebra \cite{Ozdemir:2019orp,Gomis:2019sqv}. In our approach, the NR limit of the Hietarinta CS gravity naturally extends the Bargmann algebra in a different manner, leading to the Hietarinta extended Bargmann algebra \eqref{heb}.

Considering the $\mathfrak{heb}$ algebra as a gauge algebra in replacement of the extended Bargmann algebra has interesting consequences at the dynamical level. To analyze this, let us first recall that, when constructing a CS theory for gravity, the components of the gauge field are usually identified as the dynamic variables of a geometry. However, it is in general possible to extract a geometric interpretation of a CS theory even if the field content of the topological theory exceeds the geometric one. This is the case of the Maxwell and Hietarinta gravity actions and its non-relativistic counterparts, where the components of the gauge connection that are not identified as part of the geometric variables in Cartan geometry (or Newton--Cartan in the present case) are considered additional fields of a non-geometric nature which, depending on the equations of motion they present, can propagate in such geometry. In the present case, the one-forms $\left(  \tau,e^{a},\omega,\omega^{a},m\right)  $ are identified as the geometric variables, while $\left(  s,t,k,k^{a}\right)  $ are non-geometric fields.

Let us mention that the gauge field $m$ is already present in the Bargmann algebra, whose gauging leads to Newton-Cartan theory. The role of the central charge $M$ is related with the particle' mass. There are well founded reasons for the addition of central charge gauge field $m$  ans its corresponding curvature. For instance, it allows to derive the dependent spin- and boost-connections $\omega$ and $\omega^{a}$ in terms of the independent fields $\tau,e^a$ and $m$.  On the other hand, the central charge $S$ was added to the Bargmann algebra leading to the extended Bargamann algebra, and this extension can be interpreted as the particle's spin \cite{Duval_2002, Hagen_2002}. However, there is extensive debate in the literature about this interpretation, both in general and within specific models, which we will not attempt to summarize here. An additional advantage of adding a second central extension to Galilei is that it allows to define a non-degenerate invariant tensor for the Extended Bargmann algebra. Regarding the interpretation of the central charge $T$, along its gauge field $t$, introduced in the $\mathfrak{heb}$ algebra, it is a subject of further investigation, which we hope to approach in a future work. Nevertheless, in the meantime we can say that its presence is required to define the non-degenerate invariant tensor \eqref{ITHEB}.

The non-degeneracy of the invariant tensor \eqref{ITHEB} ensures a kinematical term for each gauge field, and consequently, the dynamics of the theory is governed by the vanishing of every component of the gauge curvature in \eqref{hebeom}. These equations show that the fields $\omega$, $\tau$, $s$ and $\omega^{a}$ exhibit the same dynamical behavior than in the extended Bargmann theory. However, there are notable differences in the remaining fields $e^{a}$, $m$, $t$, $k$ and $k^{a}$. First, let us note that the fields $k$ and $k^{a}$, while exhibiting their own dynamics, appear in the equation of motion for $e^{a}$ by adding a new term to the extended Bargmann curvature $R^{a}\left(  e^{b}\right)  =de^{a}+\epsilon^{ac}\omega e_{c}+\epsilon^{ac}\tau\omega_{c}$. Whereas the extended Bargmann theory presents on-shell vanishing spatial torsion, the presence of this term modifies the vacuum of the theory, implying intrinsic non-zero spatial torsion. The same happens in the equation of motion for $m$, where the extended Bargmann curvature $R\left(m\right)  =dm+\epsilon^{ac}\omega_{a}e_{c}$ is also turned on by the $k^{a}$ gauge field. Finally, the dynamics of $t$ does not modify the geometry, since it is an auxiliary field whose presence makes it possible the formulation of a well-defined action, similar to what occurs with $s$ in the extended Bargmann context.

Let us note that the introduction of a non-vanishing torsion in a non-relativistic environment has been performed considering a completely different approach in \cite{Bergshoeff:2015ija,Bergshoeff:2017dqq,VandenBleeken:2017rij} by gauging the Schrödinger algebra, which corresponds to the conformal extension of the Bargmann algebra \cite{Bergshoeff:2014uea}. In particular, the non-vanishing torsion condition appears in the context of Lifshitz holography \cite{Christensen:2013lma} and Quantum Hall Effect \cite{Geracie:2015dea}. More recently, a non-vanishing spatial torsion in the non-relativistic realm has been encountered using the CS formalism in \cite{Concha:2021llq,Concha:2022you,Concha:2023ejs}.

Note that one can alternatively obtain the $\mathfrak{heb}$ CS gravity action \eqref{CSheb} as a non-relativistic limit of the relativistic Hietarinta CS gravity action \eqref{HCS}. To this end, the relativistic Hietarinta gauge fields has to be express as a linear combination of the non-relativistic ones as
\begin{align}
    W^{0}&=\omega+\frac{s}{2\xi^{2}}\,, & W^{a}&=\frac{\omega^{a}}{\xi}\,, & S&=\omega-\frac{s}{2\xi^{2}}\,,\notag\\ 
    E^{0}&=\xi^{2}\tau+\frac{m}{2}\,, & E^{a}&=\xi e^{a}\,, & M&=\xi^{2}\tau-\frac{m}{2}\,,\notag\\
    K^{0}&=\xi k+\frac{t}{2\xi}\,, & K^{a}&=k^{a}\,, & T&=\xi k -\frac{t}{2\xi}\,.
\end{align}
Then, the $\mathfrak{heb}$ CS gravity action is recovered from the relativistic action \eqref{HCS} after performing the limit $\xi\rightarrow\infty$.

As an ending remark, one could absorb the terms appearing in the $\alpha_2$ sector of the $\mathfrak{heb}$ gravity action \eqref{CSheb} upon the field redefinition,
\begin{align}
    e_{a}&\rightarrow e_a+\frac{\alpha_2}{\alpha_1}k_a\,, &  \tau&\rightarrow\tau+\frac{\alpha_2}{\alpha_1}k\,, & m&\rightarrow m+\frac{\alpha_2}{\alpha_1}t
\end{align}
Then, the non-relativistic CS action reads
\begin{eqnarray}
    I_{\mathfrak{heb}}&=&\int \alpha_0\left[\omega_a R^{a}\left(\omega^{b}\right)-2 s R\left(\omega\right)\right]\notag\\ &&\ +\alpha_1\left[2e_aR^{a}\left(\omega^{b}\right)-2m R\left(\omega\right)-2\tau R\left(s\right)+k_{a}R^{a}\left(k^{b}\right)-2k R\left(t\right)\right]\,.\label{CSheb2}
\end{eqnarray}
Such CS action can be seen as the non-relativistic counterpart of the Hietarinta gravity action presented in \cite{Chernyavsky:2020fqs} before including a symmetry breaking action. It would be interesting to explore if the sum of the action \eqref{CSheb2} and a symmetry breaking action contains a non-relativistic version of the Minimal Massive Gravity \cite{Bergshoeff:2014pca} as it occurs at the relativistic level in \cite{Chernyavsky:2020fqs}. One could guess that the broken non-relativistic symmetry reduces to local Nappi-Witten transformations.

\section{Non-relativistic Hietarinta Supergravity}\label{sec4}
In this section we show that a supersymmetric extension of the $\mathfrak{heb}$ algebra \eqref{heb} is achieved by applying the $S$-expansion to a $\mathcal{N}=2$ Hietarinta superalgebra following the procedure used in \cite{Concha:2019mxx,Concha:2020tqx,Concha:2021jos,Concha:2021llq}. One could apply the non-relativistic limit to the minimal Hietarinta superalgebra discussed in \cite{Concha:2023nou}. Nonetheless, the proper derivation of a non-relativistic superalgebra requires to start with a $\mathcal{N}=2$ relativistic superalgebra in order to ensure that time translational generators, in the non-relativistic realm, can be expressed as bilinear combinations of supersymmetry generators \cite{Bergshoeff:2022iyb}.

\subsection{Non-relativistic expansion of the $\mathcal{N}=2$ Hietarinta superalgebra}
Let us start with a $\mathcal{N}=2$ supersymmetric extension of the Hietarinta algebra:
\begin{align}
\left[  \tilde{J}_{A},\tilde{J}_{B}\right]   &  =\epsilon_{ABC}\tilde{J}^{C}\,,\notag\\
\left[  \tilde{J}_{A},\tilde{Z}_{B}\right]   &  =\epsilon_{ABC}\tilde{Z}^{C}\,,\notag\\
\left[  \tilde{J}_{A},\tilde{P}_{B}\right]   &  =\epsilon_{ABC}\tilde{P}^{C}\,,\notag\\
\left[  \tilde{Z}_{A},\tilde{Z}_{B}\right]   &  =\epsilon_{ABC}\tilde{P}^{C}\,,\notag\\
\left[  \tilde{J}_{A},\tilde{Q}_{\alpha}^{i}\right]   &  =-\frac{1}{2}\left(  \gamma
_{A}\right)  _{\alpha}^{\text{ \ }\beta}\tilde{Q}_{\beta}^{i}\,,\notag\\
\left[  \tilde{T},\tilde{Q}_{\alpha}^{i}\right]   &  =\frac{1}{2}\epsilon^{ij}\tilde{Q}_{\alpha}%
^{j}\,,\notag\\
\left\{  \tilde{Q}_{\alpha}^{i},\tilde{Q}_{\beta}^{j}\right\}   &  =-\delta^{ij}\left(
\gamma^{A}C\right)  _{\alpha\beta}\tilde{P}_{A}-\epsilon^{ij}C_{\alpha\beta}\tilde{S}\,, \label{SHiet}
\end{align}
which is characterized by a central charge $\tilde{S}$ and a $R$-symmetry generator $\tilde{T}$. The non-degenerate $\mathcal{N}=2$ Poincaré  superalgebra \cite{Howe:1995zm} spanned by $\{\tilde{J}_{A},\tilde{P}_{A},\tilde{Q}_{\alpha}^{i},\tilde{T},\tilde{S}\}$ is contained as a subalgebra. Let us note that the $\mathcal{N}=2$ Hietarinta superalgebra \eqref{SHiet} is quiet different from the $\mathcal{N}=2$ Maxwell superalgebra studied in \cite{Concha:2019icz} which contains an additional fermionic charge $\Sigma_{\alpha}^{i}$. The $\mathcal{N}=2$ Maxwell superalgebra has been considered as the starting relativistic superalgebra to derive the super-$\mathfrak{meb}$ one \cite{Concha:2019mxx} through the $S$-expansion method. Then, the corresponding non-relativistic expansion of the $\mathcal{N}=2$ Hietarinta superalgebra should provide us with a novel non-relativistic superalgebra different from the $\mathfrak{meb}$ one. 

The $\mathcal{N}=2$ Hietarinta superalgebra \eqref{SHiet} admits the following non-vanishing components of a non-degenerate invariant bilinear trace
\begin{align}
\left\langle \tilde{J}_{A}\tilde{J}_{B}\right\rangle  &  =\tilde{\alpha}_{0}\eta_{AB}\,, &  \left\langle \tilde{J}_{A}\tilde{P}_{B}\right\rangle &=\tilde{\alpha}_{1}\eta
_{AB}\,, \notag \\ 
\left\langle \tilde{J}_{A}\tilde{Z}_{B}\right\rangle &=\tilde{\alpha}_{2}\eta
_{AB}\,, &  
\left\langle \tilde{Z}_{A}\tilde{Z}_{B}\right\rangle & =\tilde{\alpha}_{1}\eta
_{AB}\,, \notag \\
\left\langle \tilde{T}\tilde{T}\right\rangle & =\tilde{\alpha}_{0}\,, &
 \left\langle \tilde{T}\tilde{S}\right\rangle
& =\tilde{\alpha}_{1}\,, \notag \\  
\left\langle \tilde{Q}_{\alpha}^{i}\tilde{Q}_{\beta}^{j}\right\rangle  &  =2\tilde{\alpha}_{1}%
C_{\alpha\beta}\delta^{ij}\,. \label{SInv}
\end{align}
Here, the presence of the  central charge $S$ and the $\mathfrak{so}\left(2\right)$ internal symmetry generator $T$ is required to ensure the non-degeneracy both at the relativistic level and after the non-relativistic expansion. Before applying the expansion, we first consider the decomposition of the spacetime index $A$ as in \eqref{dec} along the following subspace decomposition of the original relativistic $\mathcal{N}=2$ Hietarinta superalgebra,
\begin{align}
    V_{0}&=\{\tilde{J}_{0},\tilde{P}_{0},\tilde{Z}_{0},\tilde{T},\tilde{S},\tilde{Q}_{\alpha}^{+}\}\,, \notag\\
    V_{1}&=\{\tilde{J}_{a},\tilde{P}_{a},\tilde{Z}_{a},\tilde{Q}_{\alpha}^{-}\}\,,
\end{align}
where we have considered the following redefinition,
\begin{align}
    Q_{\alpha}^{\pm}&=\frac{1}{\sqrt{2}}\left(Q_{\alpha}^{1}\pm(\gamma^{0})_{\alpha\beta}Q_{\beta}^{2}\right)\,,
\end{align}
One can check that the subspace decomposition satisfies the following algebraic structure
\begin{align}
    \left[V_0,V_0\right]&\subset V_0\,,&  \left[V_0,V_1\right]&\subset V_1\,,&  \left[V_1,V_1\right]&\subset V_0\,. \label{subdec}
\end{align}
Let us now consider the semigroup $S_{E}^{(2)}=\{\lambda_0,\lambda_1,\lambda_2,\lambda_3\}$ as the relevant semigroup whose elements obey the following multiplication rule:
\begin{equation}
\begin{tabular}{l|llll}
$\lambda _{3}$ & $\lambda _{3}$ & $\lambda _{3}$ & $\lambda _{3}$ & $\lambda
_{3}$ \\
$\lambda _{2}$ & $\lambda _{2}$ & $\lambda _{3}$ & $\lambda _{3}$ & $\lambda
_{3}$ \\
$\lambda _{1}$ & $\lambda _{1}$ & $\lambda _{2}$ & $\lambda _{3}$ & $\lambda
_{3}$ \\
$\lambda _{0}$ & $\lambda _{0}$ & $\lambda _{1}$ & $\lambda _{2}$ & $\lambda
_{3}$ \\ \hline
& $\lambda _{0}$ & $\lambda _{1}$ & $\lambda _{2}$ & $\lambda _{3}$%
\end{tabular}
\label{ML}
\end{equation}
where $\lambda_3=0_S$ is the zero element of the semigroup satisfying $0_S\lambda_i=\lambda_i 0_S=0_S$. Let $S_{E}^{(2)}=S_0\cup S_1$ be a subset decomposition with
\begin{align}
    S_0&=\{\lambda_0,\lambda_2,\lambda_3\}\,,\notag\\
    S_1&=\{\lambda_1,\lambda_3\}\,, 
\end{align}
which is said to be resonant \cite{Izaurieta:2006zz} since it satisfies the same structure than the subspace decomposition, namely \eqref{subdec},
\begin{align}
    S_{0}\cdot S_{0}&\subset S_{0}\,, & S_{0}\cdot S_{1}&\subset S_{1}\,, & S_{1}\cdot S_{1}&\subset S_{0}\,.
\end{align}
Then, a non-relativistic superalgebra is obtained after extracting a resonant subalgebra\footnote{Following the definitions of \cite{Izaurieta:2006zz}, the expanded resonant subalgebra is given by $\mathfrak{G}=S_0\times V_0 \oplus S_1\times V_1$.} of the $S_{E}^{(2)}$-expansion of the $\mathcal{N}=2$ Hietarinta superalgebra and applying a $0_s$-reduction ($\lambda_3 T_A=0$). The expanded superalgebra is spanned by the set of generators
\begin{equation}
\{J,G_{a},S,H,P_{a},M,Z,Z_{a},T,U_1,U_2,B_1,B_2,Q_{\alpha }^{+},R_{\alpha},Q_{\alpha
}^{-}\}\,,
\end{equation}
which are related to the relativistic Hietarinta ones through the semigroup elements as
\begin{equation}
    \begin{tabular}{lll}
\multicolumn{1}{l|}{$\lambda_3$} & \multicolumn{1}{|l}{\cellcolor[gray]{0.8}} & \multicolumn{1}{|l|}{\cellcolor[gray]{0.8}} \\ \hline
\multicolumn{1}{l|}{$\lambda_2$} & \multicolumn{1}{|l}{$ S,\ \ M,\ \, T,\ \ U_2,\, B_2,\, \,R_{\alpha}$} & \multicolumn{1}{|l|}{\cellcolor[gray]{0.8}} \\ \hline
\multicolumn{1}{l|}{$\lambda_1$} & \multicolumn{1}{|l}{\cellcolor[gray]{0.8}} & \multicolumn{1}{|l|}{$G_a,\ P_a,\ Z_a, \  Q^{-}_{\alpha}$} \\ \hline
\multicolumn{1}{l|}{$\lambda_0$} & \multicolumn{1}{|l}{$ J,\ \ H,\ \ Z,\ \ U_1,\, B_1,\, \,Q^{+}_{\alpha}$} & \multicolumn{1}{|l|}{\cellcolor[gray]{0.8}} \\ \hline
\multicolumn{1}{l|}{} & \multicolumn{1}{|l}{$\tilde{J}_0,\ \tilde{P}_0,\ \tilde{Z}_0,\ \, \tilde{T},\ \, \tilde{S}, \ \  \tilde{Q}^{+}_{\alpha}$} & \multicolumn{1}{|l|}{$\tilde{J}_{a},\ \, \tilde{P}_{a},\ \, \tilde{Z}_{a}, \ \tilde{Q}^{-}_{\alpha}$} 
\end{tabular}\label{Sexp}%
\end{equation}
At the bosonic level, the expanded generators satisfy the $\mathfrak{heb}$ algebra \eqref{heb}. On the other hand, the ferminionic content satisfy the following non-vanishing (anti-)commutation relations:
\begin{align}
    \left[  J,Q_{\alpha}^{\pm}\right]   &  =-\frac{1}{2}(\gamma_0)_{\alpha }^{\ \beta }Q_{\beta}^{\pm}\,, & \left[  G_a,Q_{\alpha}^{+}\right]   &  =-\frac{1}{2}(\gamma_0)_{\alpha }^{\ \beta }Q_{\beta}^{-}\,,  \notag \\
\left[  S,Q_{\alpha}^{+}\right]   &  =-\frac{1}{2}(\gamma_0)_{\alpha }^{\ \beta }R_{\beta}\,, & \left[  G_a,Q_{\alpha}^{-}\right]   &  =-\frac{1}{2}(\gamma_0)_{\alpha }^{\ \beta }R_{\beta}\,,
\notag \\
\left[  J,R_{\alpha}\right]   &  =-\frac{1}{2}(\gamma_0)_{\alpha }^{\ \beta }R_{\beta}\,,   &
\left[  U_1,Q_{\alpha}^{\pm}\right]   &  =\pm\frac{1}{2}(\gamma_0)_{\alpha }^{\ \beta }Q_{\beta}^{\pm}\,, \notag \\
\left[  U_1,R_{\alpha}\right]   &  =\frac{1}{2}(\gamma_0)_{\alpha }^{\ \beta }R_{\beta}\,, & \left[  U_2,Q_{\alpha}^{+}\right]   &  =-\frac{1}{2}(\gamma_0)_{\alpha }^{\ \beta }R_{\beta}\,, \notag \\
\left\{  Q_{\alpha}^{+},Q_{\beta}^{+}\right\}   &  =-\left(
\gamma^{0}C\right)_{\alpha\beta}H-\left(\gamma^{0}C\right)_{\alpha\beta}B_1\,, \notag \\
\left\{  Q_{\alpha}^{+},Q_{\beta}^{-}\right\}   &  =-\left(
\gamma^{a}C\right)_{\alpha\beta}P_{a}\,, \notag \\
\left\{  Q_{\alpha}^{-},Q_{\beta}^{-}\right\}   &  =-\left(
\gamma^{0}C\right)_{\alpha\beta}M-\left(\gamma^{0}C\right)_{\alpha\beta}B_2\,, \notag \\
\left\{  Q_{\alpha}^{+},R_{\beta}\right\}   &  =-\left(
\gamma^{0}C\right)_{\alpha\beta}M-\left(\gamma^{0}C\right)_{\alpha\beta}B_2\,. \label{sheb} 
\end{align}
The expanded superalgebra corresponds to a supersymmetric extension of the $\mathfrak{heb}$ algebra \eqref{heb} and can be seen as an extension of the so-called extended Bargmann superalgebra \cite{Bergshoeff:2016lwr} spanned by $\{J,G_a,H,P_a,S,M,Q_{\alpha}^{\pm},R_{\alpha}\}$. As in the extended super-Bargmann case, the non-relativistic superalgebra contains a third Majorana fermionic charge $R_{\alpha}$ which, along the central charges $\{B_1,B_2\}$ and the additional bosonic generators $\{U_1,U_2\}$, ensures the non-degeneracy of the invariant bilinear trace. The presence of the $U_1$ and $U_2$ generators, which appear as expansion of the $R$-symmetry generators of the $\mathcal{N}=2$ Hietarinta superalgebra, is not new in the literature and also appears in the Newton-Hooke superalgebra introduced in \cite{Ozdemir:2019tby}. 

Unlike its bosonic counterpart, the $\mathfrak{heb}$ superalgebra is not isomorphic to the Maxwellian extended Bargmann one presented in \cite{Concha:2019mxx}. The $\mathfrak{meb}$ superalgebra is characterized by six different fermionic charges $\{Q_{\alpha}^{\pm},R_{\alpha},\Sigma_{\alpha}^{\pm},W_{\alpha}\}$ and corresponds to the non-relativistic version of the $\mathcal{N}=2$ Maxwell superalgebra \cite{Concha:2019icz}. The interchanging of the bosonic generators $P_a\leftrightarrow Z_a, H\leftrightarrow Z, M\leftrightarrow T$ in the $\mathfrak{heb}$ superalgebra \eqref{heb} and \eqref{sheb} reproduces a different Maxwellian generalization of the extended Bargmann superalgebra which can be seen as the non-relativistic version of a $\mathcal{N}=2$ non-standard Maxwell superalgebra \cite{Lukierski:2010dy,Concha:2020atg}. Such non-standard $\mathfrak{meb}$ superalgebra is not a good candidate to describe non-relativistic supergravity since the time translational generators $P_a$ are no longer expressed as bilinear combinations of supersymmetry generators. Nevertheless, the Hietarinta basis does not suffer from such peculiarity and, as we shall see, can offer a truly extension of the extended Bargmann supergravity.

A limiting process can also be performed to derive the $\mathfrak{heb}$ superalgebra from three copies of the Nappi-Witten algebra, one of which is augmented by supersymmetry \cite{Concha:2020tqx,Concha:2020eam},
\begin{align}
\left[  \hat{J},\hat{P}_{a}\right]   &  =\epsilon_{ab}\hat{P}_{b}\,, &\left[\hat{P}_{a},\hat{P}_{b} \right]&=-\epsilon_{ab} \hat{T}\,, \notag\\
\left[ \hat{J},\hat{\mathcal{Q}}_{\alpha}^{\pm}\right] & = -\frac{1}{2}(\gamma_0)_{\alpha }^{\ \beta }\hat{\mathcal{Q}}_{\beta}^{\pm}\,, & \left[  \hat{P}_a,\hat{\mathcal{Q}}_{\alpha}^{+}\right]   &  =-\frac{1}{2}(\gamma_0)_{\alpha }^{\ \beta }\hat{\mathcal{Q}}_{\beta}^{-}\,,  \notag \\
\left[  \hat{T},\hat{\mathcal{Q}}_{\alpha}^{+}\right]   &  =-\frac{1}{2}(\gamma_0)_{\alpha }^{\ \beta }\hat{\mathcal{R}}_{\beta}\,, & \left[  \hat{P}_a,\hat{\mathcal{Q}}_{\alpha}^{-}\right]   &  =-\frac{1}{2}(\gamma_0)_{\alpha }^{\ \beta }\hat{\mathcal{R}}_{\beta}\,,
\notag \\
\left[  \hat{J},\hat{\mathcal{R}}_{\alpha}\right]   &  =-\frac{1}{2}(\gamma_0)_{\alpha }^{\ \beta }\hat{\mathcal{R}}_{\beta}\,,   &
\left[  \hat{Y}_1,\hat{\mathcal{Q}}_{\alpha}^{\pm}\right]   &  =\pm\frac{1}{2}(\gamma_0)_{\alpha }^{\ \beta }\hat{\mathcal{Q}}_{\beta}^{\pm}\,, \notag \\
\left[  \hat{Y}_1,\hat{\mathcal{R}}_{\alpha}\right]   &  =\frac{1}{2}(\gamma_0)_{\alpha }^{\ \beta }\hat{\mathcal{R}}_{\beta}\,, & \left[  \hat{Y}_2,\hat{\mathcal{Q}}_{\alpha}^{+}\right]   &  =-\frac{1}{2}(\gamma_0)_{\alpha }^{\ \beta }\hat{\mathcal{R}}_{\beta}\,, \notag \\
\left\{  \hat{\mathcal{Q}}_{\alpha}^{+},\hat{\mathcal{Q}}_{\beta}^{+}\right\}   &  =-\left(
\gamma^{0}C\right)_{\alpha\beta}\hat{J}-\left(\gamma^{0}C\right)_{\alpha\beta}\hat{Y}_1\,, \notag \\
\left\{  \hat{\mathcal{Q}}_{\alpha}^{+},\hat{\mathcal{Q}}_{\beta}^{-}\right\}   &  =-\left(
\gamma^{a}C\right)_{\alpha\beta}\hat{P}_{a}\,, \notag \\
\left\{  \hat{\mathcal{Q}}_{\alpha}^{-}\hat{\mathcal{Q}}_{\beta}^{-}\right\}   &  =-\left(
\gamma^{0}C\right)_{\alpha\beta}\hat{T}-\left(\gamma^{0}C\right)_{\alpha\beta}\hat{Y}_2\,, \notag \\
\left\{  \hat{\mathcal{Q}}_{\alpha}^{+},\hat{\mathcal{R}}_{\beta}\right\}   &  =-\left(
\gamma^{0}C\right)_{\alpha\beta}\hat{T}-\left(\gamma^{0}C\right)_{\alpha\beta}\hat{Y}_2\,. \label{sNW} 
\end{align}
Indeed the $\mathfrak{heb}$ superalgebra \eqref{heb} and \eqref{sheb} is obtained by performing the flat limit $\ell\rightarrow\infty$ after the following redefinition of the generators,
\begin{align}
    G_{a}&=\bar{P}_{a}+\hat{P}_{a}+P_{a}^{*}\,, & Z_{a}&=\frac{1}{\ell}\left(\hat{P}_{a}-P_{a}^{*}\right)\,,  & P_{a}&=\frac{1}{\ell^{2}}\left(\hat{P}_{a}+P_{a}^{*}\right)\,, \notag \\
    J&=\bar{J}+\hat{J}+J^{*}\,, & Z&=\frac{1}{\ell}\left(\hat{J}-J^{*}\right)\,,  & H&=\frac{1}{\ell^{2}}\left(\hat{J}+J^{*}\right)\,, \notag \\
     S&=\bar{T}+\hat{T}+T^{*}\,, & T&=\frac{1}{\ell}\left(\hat{T}-T^{*}\right)\,,  & M&=\frac{1}{\ell^{2}}\left(\hat{T}+T^{*}\right)\,, \notag \\
     Q_{\alpha}^{+}&=\frac{\sqrt{2}}{\ell}\hat{\mathcal{Q}}_{\alpha}^{+}\,, &  Q_{\alpha}^{-}&=\frac{\sqrt{2}}{\ell}\hat{\mathcal{Q}}_{\alpha}^{-}\,, & R_{\alpha}&=\frac{\sqrt{2}}{\ell}\hat{\mathcal{R}}_{\alpha}\,, \notag \\
     U_{1}&=\hat{Y}_{1}+Y_{1}^{*}\,, & B_{1}&=\frac{1}{\ell^{2}}\hat{Y}_{1}\,, \notag \\
     U_{2}&=\hat{Y}_{2}+Y_{2}^{*}\,, & B_{2}&=\frac{1}{\ell^{2}}\hat{Y}_{2}\,. \label{redef}
\end{align} 
Here the subset $\{J^{*},P_{a}^{*},T^{*},Y_{1}^{*},Y_{2}^{*}\}$ defines a Nappi-Witten algebra coupled to two $\mathfrak{u}(1)$ generators. The subset $\{\bar{J},\bar{P}_{a},\bar{T}\}$ is the standard Nappi-Witten algebra \cite{Nappi:1993ie,Figueroa-OFarrill:1999cmq}, while its supersymmetric extension spanned by $\{\hat{J},\hat{P}_{a},\hat{T},\hat{\mathcal{Q}}_{\alpha}^{+},\hat{\mathcal{Q}}_{\alpha}^{-},\hat{\mathcal{R}}_{\alpha},\hat{Y}_{1},\hat{Y}_{2}\}$ satisfies the (anti-)commutators given in \eqref{sNW}. Otherwise, the $\mathfrak{heb}$ superalgebra \eqref{heb} and \eqref{sheb} can also be obtained from the direct sum of the extended Newton-Hooke superalgebra \cite{Ozdemir:2019tby} and the Nappi-Witten algebra endowed with two $\mathfrak{u}\left(1\right)$ generators. Indeed the super-$\mathfrak{heb}$ (anti-)commutation relations appear after considering the following redefinition,
\begin{align}
G_a&=P_{a}^{\star}+\texttt{G}_{a}\,, & P_a&=\frac{1}{\ell^{2}}\texttt{G}_{a}\,, & Z_{a}&=\texttt{P}_{a}\,, \notag \\
J&=J^{\star}+\texttt{J}\,, & H&=\frac{1}{\ell^{2}}\texttt{J}\,, & Z&=\texttt{H}\,, \notag \\
S&=T^{\star}+\texttt{S}\,, & M&=\frac{1}{\ell^{2}}\texttt{S}\,, & T&=\texttt{M}\,, \notag \\
Q_{\alpha}^{+}&=\sqrt{\frac{1}{\ell}}\texttt{Q}_{\alpha}^{+}\,, &  Q_{\alpha}^{-}&=\sqrt{\frac{1}{\ell}}\texttt{Q}_{\alpha}^{-}\,, & R_{\alpha}&=\sqrt{\frac{1}{\ell}}\texttt{R}_{\alpha}\,, \notag \\
U_{1}&=Y_{1}^{\star}+\texttt{Y}_{1}\,, & B_{1}&=\frac{1}{\ell^{2}}\texttt{Y}_{1}\,, \notag \\
     U_{2}&=Y_{2}^{\star}+\texttt{Y}_{2}\,, & B_{2}&=\frac{1}{\ell^{2}}\texttt{Y}_{2}\,, \label{redef2}
\end{align}
and performing the limit $\ell\rightarrow\infty$. Here the subset of generators $\{\texttt{J},\texttt{G}_{a},\texttt{H},\texttt{P}_{a},\texttt{S},\texttt{M},\texttt{Y}_{1},\texttt{Y}_{2},\texttt{Q}^{+}_{\alpha},\texttt{Q}^{-}_{\alpha},\texttt{R}_{\alpha}\}$ satisfies the extended Newton-Hooke superalgebra \cite{Ozdemir:2019tby}, while $\{J^{\star},P_{a}^{\star},T^{\star},Y_{1}^{\star},Y_{2}^{\star}\}$ defines the Nappi-Witten algebra \cite{Nappi:1993ie,Figueroa-OFarrill:1999cmq} endowed with two $\mathfrak{u}(1)$ generators.

\subsection{Three-dimensional Hietarinta extended Bargmann supergravity action}
In this section we present a novel three-dimensional non-relativistic supergravity theory without cosmological constant that extends the extended Bargmann supergravity of \cite{Bergshoeff:2016lwr}. First, let us consider the gauge connection one-form $A$ for the $\mathfrak{heb}$ superalgebra \eqref{heb} and $\eqref{sheb}$:
\begin{eqnarray}
    A&=&\omega J +\omega_a G_a + \tau H + e^{a}P_{a} + k Z + k^{a} Z_{a}+s S +m M+ t T\notag \\
    & &+u_1 U_1+ u_2 U_2 + b_1 B_1 + b_2 B_2 + \bar{\psi}^{+} Q^{+} + \bar{\psi}^{-} Q^{-} +\bar{\rho} \,R \,. \label{Asheb}
\end{eqnarray}
In addition to the bosonic $\mathfrak{heb}$ content, the proper supersymmetric extension requires the presence of the gauge fields $u_1$, $u_2$, $b_1$ and $b_2$. The corresponding curvature two-form reads
\begin{eqnarray}
    F&=&F\left(\omega\right)J+F^{a}\left(\omega^{b}\right)G_a+F\left(\tau\right)H+F^{a}\left(e^{b}\right)P_a+F\left(s\right)S+F\left(m\right)M+F\left(t\right)T\notag\\
    &&+F\left(k\right)Z+F^{a}\left(k^{b}\right)Z_{a}+F\left(u_1\right)U_1+F\left(u_2\right)U_2+F\left(b_1\right)B_1+F\left(b_2\right)B_2\notag\\
    &&+\nabla \bar{\psi}^{+} Q^{+}+\nabla \bar{\psi}^{-} Q^{-} +\nabla \bar{\rho}\,R\,. \label{scurv}
\end{eqnarray}
Here, the bosonic curvature two-forms are given by
\begin{align}
F\left(\omega\right)&=\texttt{R}\left(\omega\right)\,,&
F^{a}\left(\omega^{b}\right)&=\texttt{R}^{a}\left(\omega^{b}\right)\,,\notag\\
F\left(\tau\right)&=\texttt{R}\left(\tau\right)+\frac{1}{2}\bar{\psi}^{+}\gamma^{0}\psi^{+}\,,&
F^{a}\left(e^{b}\right)&=\texttt{R}^{a}\left(e^{b}\right)+\bar{\psi}^{+}\gamma^{a}\psi^{-}\,,\notag\\
F\left(s\right)&=\texttt{R}\left(s\right)\,,&
F\left(m\right)&=\texttt{R}\left(m\right)+\frac{1}{2}\bar{\psi}^{-}\gamma^{0}\psi^{-}+\bar{\psi}^{+}\gamma^{0}\rho\,,\notag\\
F\left(k\right)&=\texttt{R}\left(k\right)\,,&
F^{a}\left(k^{b}\right)&=\texttt{R}^{a}\left(k^{b}\right)\,,\notag\\
F\left(t\right)&=\texttt{R}\left(t\right)\,, & F\left(u_1\right)&=du_1\,, \notag\\
F\left(u_2\right)&=du_2\,, & F\left(b_1\right)&=db_1+\frac{1}{2}\bar{\psi}^{+}\gamma^{0}\psi^{+}\,, \notag\\
F\left(b_2\right)&=db_2+\frac{1}{2}\bar{\psi}^{-}\gamma^{0}\psi^{-}+\bar{\psi}^{+}\gamma^{0}\rho\,, \label{boscurv}
\end{align}
where $\texttt{R}\left(\omega\right)$, $\texttt{R}^{a}\left(\omega^{b}\right)$, $\texttt{R}\left(\tau\right)$, $\texttt{R}^{a}\left(e^{b}\right)$, $\texttt{R}\left(s\right)$, $\texttt{R}\left(m\right)$, $\texttt{R}\left(k\right)$, $\texttt{R}^{a}\left(k^{b}\right)$ and $\texttt{R}\left(t\right)$ are the $\mathfrak{heb}$ curvatures defined in \eqref{acurv} and \eqref{hebeom}. On the other hand, the covariant derivatives of the spinor 1-forms read
\begin{eqnarray}
    \nabla \psi^{+}&=&d\psi^{+}+\frac{1}{2}\omega\gamma_0\psi^{+}-\frac{1}{2}u_1\gamma_0 \psi^{+}\,,\notag\\
    \nabla \psi^{-}&=&d\psi^{-}+\frac{1}{2}\omega\gamma_0\psi^{-}+\frac{1}{2}\omega^{a}\gamma_a \psi^{+}+\frac{1}{2}u_1\gamma_0 \psi^{-}\,,\notag\\
    \nabla \rho&=&d\rho+\frac{1}{2}\omega\gamma_0\rho+\frac{1}{2}\omega^{a}\gamma_a \psi^{-}+\frac{1}{2}s\gamma_0\psi^{+}-\frac{1}{2}u_1\gamma_0 \rho+\frac{1}{2}u_2\gamma_0 \psi^{+}\,. \label{fermcurv}
\end{eqnarray}
Notice that when the Hietarinta gauge fields vanish  ($k=k_a=0$) and no additional bosonic content is present ($u_1=u_2=b_1=b_2=0$), the extended Bargmann curvatures \cite{Bergshoeff:2016lwr} are recovered. 

The presence of central charges and $R$-symmetry generators is not random but ensures the non-degeneracy of the invariant bilinear trace. Indeed the $\mathfrak{heb}$ superalgebra, given by \eqref{heb} and \eqref{sheb}, admits the following non-vanishing components of a non-degenerate invariant tensor:
\begin{align}
    \langle J S \rangle&=-\alpha_0\,, & \langle G_{a}G_{b}\rangle &=\alpha_0 \delta_{ab}\,, & \langle J M \rangle &=-\alpha_1\,,\notag \\
    \langle H S \rangle &=-\alpha_1 \,, & \langle G_{a} P_{b} \rangle &=\alpha_1 \delta_{ab}\,, & \langle T Z \rangle &=-\alpha_1 \,,\notag \\
    \langle J T \rangle &=-\alpha_2 \,, & \langle Z_{a}Z_{b} \rangle&=\alpha_1 \delta_{ab}\,, &  \langle U_1 U_2 \rangle & =\alpha_0  \notag \\  \langle Z S \rangle &=-\alpha_2\,, & \langle G_{a}Z_{b}\rangle &=\alpha_2\delta_{ab}\,, & \langle U_1 B_2 \rangle &= \alpha_1 \,, \notag \\
    \langle U_2 B_1 \rangle &= \alpha_1 \,, & \langle Q_{\alpha}^{-} Q_{\beta}^{-} \rangle &= 2\alpha_1 C_{\alpha\beta}\,, & \langle Q_{\alpha}^{+} R_{\beta} \rangle &= 2\alpha_1 C_{\alpha\beta}\,.  \label{ITSHEB}
\end{align}
The invariant tensor is obtained from the relativistic $\mathcal{N}=2$ Hietarinta one \eqref{SInv} by expressing the non-relativistic generators in terms of the relativistic ones as in \eqref{Sexp} and considering the multiplication law \eqref{ML} of the semigroup $S_{E}^{\left(2\right)}$. Here, $\alpha_0$, $\alpha_1$ and $\alpha_2$ are related to the relativistic constants though the $\lambda_2$ semigroup element as follows,
\begin{align}
    \alpha_0&=\lambda_2 \tilde{\alpha}_0\,, & \alpha_1&=\lambda_2 \tilde{\alpha}_1\,, & \alpha_2&=\lambda_2 \tilde{\alpha}_2\,.\label{expp}
\end{align}
Let us note that the non-degeneracy is guaranteed for $\alpha_1\neq 0$. Given that the terms along $\alpha_0$ and $\alpha_2$ are already determined in the $\mathfrak{heb}$ gravity framework \eqref{CSheb} (plus contributions of $u_1$ and $u_2$), we will focus on the supersymmetric sector that is proportional to $\alpha_1$. Therefore, for the construction of the non-relativistic supergravity action, we can, without loss of generality, set $\alpha_0=\alpha_2=0$ as this does not affect non-degeneracy.

Then, the CS supergravity action for the $\mathfrak{heb}$ superalgebra \eqref{heb} and \eqref{sheb} is obtained considering the gauge connection one-form \eqref{Asheb} and the non-vanishing components of the invariant tensor \eqref{ITSHEB} into the general expression of the CS action \eqref{CSaction},
\begin{eqnarray}
    I_{s\mathfrak{heb}}&=&\int  2e_aR^{a}\left(\omega^{b}\right)-2m R\left(\omega\right)-2\tau R\left(s\right)+k_{a}R^{a}\left(k^{b}\right)-2k R\left(t\right)+2u_1db_2+2u_2db_1\notag\\
    &&-2\bar{\psi}^{-}\nabla\psi^{-}-2\bar{\psi}^{+}\nabla\rho-2\bar{\rho}\nabla\psi^{+}\,,\label{CSsheb}
\end{eqnarray}
where we have set $\alpha_1=1$. Here $R\left(\omega\right)$, $R^{a}\left(\omega^{b}\right)$, $R\left(s\right)$, $R^{a}\left(k^{b}\right)$ and $R\left(t\right)$ are the $\mathfrak{heb}$ curvatures defined in \eqref{acurv}. The non-relativistic supergravity action derived here differs significantly from the Newton-Cartan supergravity theory \cite{Andringa:2013mma,Bergshoeff:2015ija} and the extended Bargmann one \cite{Bergshoeff:2016lwr}. 

Let us recall that the field content of extended Bargmann supergravity is
composed of the one-forms $\omega$, $\omega^{b}$, $\tau$, $e^{b}$, $m$, $s$,
$\psi^{+}$, $\psi^{-}$ and $\rho$, where, in particular, the equation of motion for
$e^{a}$ sets on-shell vanishing spatial supertorsion. In the $\mathfrak{heb}$
supergravity theory, these fields are also present and most of them exhibit
the same dynamical behavior, except for $e^{a}$ and $m$. In fact, in addition
to the field content of extended Bargmann supergravity, this theory includes
the one-forms $k$, $k^{a}$, $t$, $u_{1}$, $u_{2}$, $b_{1}$ and $b_{2}$ as
dynamical fields. Similar to what we found in the bosonic case, the first two
gauge fields $k$ and $k^{a}$ change the vacuum of the resulting supergravity
theory by imposing non-vanishing on-shell spatial super-torsion as a
consequence of the equation of motion for $e^{a}$. Moreover, the equation of
motion for $m$ is also modified, implying that its extended Bargmann
super-curvature is also non-vanishing on-shell. Thus, the dynamics of extended
Bargmann supergravity is modified in these two geometric aspects, while the
equations of motion of the remaining gauge fields do not affect the resulting
allowed Newton--Cartan supergeometries, but their presence guarantees a
well-defined action principle, and consequently, the validity of the equations
of motion. On the other hand, despite the isomorphism existing between the
$\mathfrak{heb}$ and the Maxwellian extended Bargmann algebra
\cite{Aviles:2018jzw}, the action \eqref{CSsheb}, when supersymmetry is
incorporated, differs fundamentally from the $\mathfrak{meb}$ supergravity
action \cite{Concha:2019mxx}. As it was described in the context of the
extended Bargmann supergravity, the main difference between this theory and
the $\mathfrak{meb}$ supergravity is the spatial component of the
super-torsion, which is not zero but is instead proportional to the additional
gauge fields $k^{a}$ and $k$,
\begin{eqnarray}
    \texttt{T}^{a}\left(e^{b}\right)&=&-\epsilon^{ac}kk_c\,, \notag
\end{eqnarray}
where $\texttt{T}^{a}\left(e^{b}\right)=de^{a}+\epsilon^{ac}\omega e_{c}+\epsilon^{ac}\tau\omega_{c}+\bar{\psi}^{+}\gamma^{a}\psi^{-}$. 

The $\mathfrak{heb}$ supergravity action can alternatively be obtained by expanding the relativistic $\mathcal{N}=2$ Hietarianta supergravity action,
\begin{eqnarray}
I_{s\mathcal{H}} &=&\frac{k}{4\pi }\int \left( 2E^{A}R_{A}+K^{A}\mathcal{F}_{A}+\texttt{t}d\texttt{s}-2\bar{\Psi}^{i}\nabla \Psi^{i} \right) \,,  \label{sHCS}
\end{eqnarray}
where $R^{A}$ and $\mathcal{F}^{A}$ are given in \eqref{bosc} and
\begin{eqnarray}
    \nabla\Psi^{i}=d\Psi^{i}+\frac{1}{2}\omega^{a}\Gamma_{a}\Psi^{i}+\texttt{t}\epsilon^{ij}\Psi^{j}\,.
\end{eqnarray}
Let us note that the non-relativistic gauge fields can be expressed as a resonant expansion of the relativistic ones as
\begin{align}
    \omega&=\lambda_0 W^{0}\,, & \omega^{a}&=\lambda_1 W^{a}\,, & s&=\lambda_2 W^{0}\,, \notag \\
    \tau&=\lambda_0 E^{0}\,, & e^{a}&=\lambda_1 E^{a}\,, & m&=\lambda_2 E^{0}\,, \notag\\
    k&=\lambda_0 K^{0}\,, & k^{a}&=\lambda_1 K^{a}\,, & t&=\lambda_2 K^{0}\,, \notag \\
    u_1&=\lambda_0 \texttt{t}\,, & u_2&=\lambda_2 \texttt{t}\,, & b_1&=\lambda_0 \texttt{b}\,, \notag \\
    b_2&= \lambda_2 \texttt{b}\,, & \psi_{\alpha}^{+}&=\lambda_0 \Psi_{\alpha}^{+}\,,  & \psi_{\alpha}^{-}&=\lambda_1 \Psi_{\alpha}^{}\,, \notag\\
    \rho_{\alpha}&=\lambda_2 \Psi_{\alpha}^{+}\,,\label{expgf}
\end{align}
whose semigroup elements belong to $S_{E}^{\left(2\right)}$ \eqref{ML}. Here, we have considered the following redefinition,
\begin{eqnarray}
    \Psi^{\pm}&=&\frac{1}{\sqrt{2}}\left(\Psi_{\alpha}^{1}\pm\left(\gamma^{0}\right)_{\alpha\beta}\Psi_{\beta}^{2}\right)\,.
\end{eqnarray}
The derivation of the $\mathfrak{heb}$ supergravity action \eqref{CSsheb} is obtained considering the expanded gauge fields \eqref{expgf} and the expanded parameters \eqref{expp} along the properties of the $0_S$-reduction.

\subsection{The inclusion of a cosmological constant term}
A cosmological constant term can be accommodated into the Hietarinta extended Bargmann supergravity. We have previously shown that the $\mathfrak{heb}$ superalgebra can alternatively be derived from three copies of the Nappi-Witten algebra, one of which is augmented by supersymmetry. One can show that the generators defined in terms of the Nappi-Witten ones in \eqref{redef} satisfy the $\mathfrak{heb}$ commutation relations \eqref{heb} and those appearing in \eqref{sheb} along the following ones:
\begin{align}
\left[  Z,P_{a}\right]   &  =\frac{1}{\ell^2}\epsilon_{ab}Z_{b}\,, & \left[  P_{a},P_{b}\right]   &  =-\frac{1}{\ell^2}\epsilon_{ab}M\,,&
\left[  H,Z_{a}\right]   &  =\frac{1}{\ell^2}\epsilon_{ab}Z_{b}\,,\notag \\
\left[  H,P_{a}\right]   &  =\frac{1}{\ell^2}\epsilon_{ab}P_{b}\,, & \left[  Z_{a},P_{b}\right]   &  =-\frac{1}{\ell^2}\epsilon_{ab}T\,,& \left[  Z,Q_{\alpha}^{\pm}\right]   &  =-\frac{1}{2\ell}(\gamma_0)_{\alpha }^{\ \beta }Q_{\beta}^{\pm}\,,
\notag \\
\left[  Z,R_{\alpha}\right]   &  =-\frac{1}{2\ell}(\gamma_0)_{\alpha }^{\ \beta }R_{\beta}\,, & \left[ H,Q_{\alpha}^{\pm}\right]   &  =-\frac{1}{2\ell^2}(\gamma_0)_{\alpha }^{\ \beta }Q_{\beta}^{\pm}\,, & \left[  H,R_{\alpha}\right]   &  =-\frac{1}{2\ell^2}(\gamma_0)_{\alpha }^{\ \beta }R_{\beta}\,, \notag\\
\left[  T,Q_{\alpha}^{+}\right]   &  =-\frac{1}{2\ell}(\gamma_0)_{\alpha }^{\ \beta }R_{\beta}\,, & \left[  M,Q_{\alpha}^{+}\right]   &  =-\frac{1}{2\ell^2}(\gamma_0)_{\alpha }^{\ \beta }R_{\beta}\,, & \left[  Z_a,Q_{\alpha}^{+}\right]   &  =-\frac{1}{2\ell}(\gamma_a)_{\alpha }^{\ \beta }Q_{\beta}^{-}\,,  \notag \\
\left[  Z_a,Q_{\alpha}^{-}\right]   &  =-\frac{1}{2\ell}(\gamma_a)_{\alpha }^{\ \beta }R_{\beta}\,, & \left[  P_a,Q_{\alpha}^{+}\right]   &  =-\frac{1}{2\ell^2}(\gamma_a)_{\alpha }^{\ \beta }Q_{\beta}^{-}\,, & \left[  P_a,Q_{\alpha}^{-}\right]   &  =-\frac{1}{2\ell^2}(\gamma_a)_{\alpha }^{\ \beta }R_{\beta}\,, \notag\\
\left[  B_1,Q_{\alpha}^{\pm}\right]   &  =\pm\frac{1}{2\ell^2}(\gamma_0)_{\alpha }^{\ \beta }Q_{\beta}^{\pm}\,, &
\left[  B_1,R_{\alpha}\right]   &  =\frac{1}{2\ell^2}(\gamma_0)_{\alpha }^{\ \beta }R_{\beta}\,, & \left[  B_2,Q_{\alpha}^{+}\right]   &  =-\frac{1}{2\ell^2}(\gamma_0)_{\alpha }^{\ \beta }R_{\beta}\,, \label{eheb}
\end{align}
where $\ell$ is the length parameter which is related to the cosmological constant $\Lambda$ through $\Lambda\propto \frac{1}{\ell^2}$. On the other hand, the fermionic generators satisfy the following anti-commutators:
\begin{align}
  \left\{  Q_{\alpha}^{+},Q_{\beta}^{+}\right\}   &  =-\left(
\gamma^{0}C\right)_{\alpha\beta}H-\frac{1}{\ell}\left(
\gamma^{0}C\right)_{\alpha\beta}Z-\left(\gamma^{0}C\right)_{\alpha\beta}B_1\,, \notag \\
\left\{  Q_{\alpha}^{+},Q_{\beta}^{-}\right\}   &  =-\left(
\gamma^{a}C\right)_{\alpha\beta}P_{a}-\frac{1}{\ell}\left(
\gamma^{a}C\right)_{\alpha\beta}Z_{a}\,, \notag \\
\left\{  Q_{\alpha}^{-},Q_{\beta}^{-}\right\}   &  =-\left(
\gamma^{0}C\right)_{\alpha\beta}M-\frac{1}{\ell}\left(
\gamma^{0}C\right)_{\alpha\beta}T-\left(\gamma^{0}C\right)_{\alpha\beta}B_2\,, \notag \\
\left\{  Q_{\alpha}^{+},R_{\beta}\right\}   &  =-\left(
\gamma^{0}C\right)_{\alpha\beta}M-\frac{1}{\ell}\left(
\gamma^{0}C\right)_{\alpha\beta}T-\left(\gamma^{0}C\right)_{\alpha\beta}B_2\,. \label{esheb} 
\end{align}
The superalgebra given by \eqref{heb},\eqref{sheb}, \eqref{eheb} and \eqref{esheb} can be seen as an enlargement of the Hietarinta extended Bargmann superalgebra and can be written as the direct sum of the extended Newton-Hooke superalgebra \cite{Ozdemir:2019tby} and the Nappi-Witten algebra \cite{Nappi:1993ie, Figueroa-OFarrill:1999cmq}. One can check that the cosmological extension of the $\mathfrak{heb}$ superalgebra, which we shall denote as $\overline{\mathfrak{heb}}$ superalgebra, can alternatively be recovered as a non-relativistic expansion of the so-called $\mathcal{N}=2$ non-standard AdS-Lorentz superalgebra \cite{Concha:2018jxx, Concha:2020atg}. Naturally, in the vanishing cosmological constant limit $\ell\rightarrow\infty$, the non-relativistic superalgebra reduces to the $\mathfrak{heb}$ superalgebra \eqref{heb} and \eqref{sheb}. In particular, the $B_1$ and $B_2$ generators become central charges in the flat limit. 

The $\overline{\mathfrak{heb}}$ superalgebra admits a non-degenerate invariant tensor whose non-vanishing components are given by the bosonic $\mathfrak{heb}$ ones appearing in \eqref{ITSHEB} along
\begin{align}
    \langle H T \rangle&=-\frac{\alpha_2}{\ell^2}\,, & \langle P_{a}P_{b}\rangle &=\frac{\alpha_1}{\ell^2} \delta_{ab}\,, \notag \\
    \langle Z M \rangle &=-\frac{\alpha_2}{\ell^2}\,, & \langle P_{a}Z_{b}\rangle &=\frac{\alpha_2}{\ell^2} \delta_{ab}\,,\notag \\
    \langle H M \rangle&=-\frac{\alpha_1}{\ell^2}\,, &  \langle B_1 B_2 \rangle&=\frac{\alpha_2}{\ell^2}\,,\notag\\
    \langle Q_{\alpha}^{-} Q_{\beta}^{-} \rangle&=2\left(\alpha_1+\frac{\alpha_2}{\ell}\right)C_{\alpha\beta}\,, & \langle Q_{\alpha}^{+} R_{\beta} \rangle &= 2\left(\alpha_1+\frac{\alpha_2}{\ell}\right) C_{\alpha\beta}\,. \label{ITESHEB}
\end{align}
Let us note that $\alpha_0$, $\alpha_1$ and $\alpha_2$ can be written in terms of the constants $\mu$, $\nu$ and $\sigma$ appearing in the three copies of the Nappi-Witten algebra, one of which is augmented by supersymmetry. In particular, we have
\begin{align}
    \alpha_0&=\mu+\nu+\sigma\,, & \alpha_1&=\frac{1}{\ell}\left(\nu+\sigma\right) & \alpha_2&=\frac{1}{\ell^2}\left(\nu-\sigma\right)\,, \label{redef3}
\end{align}
where $\nu$ is related to the invariant tensor of the Nappi-Witten superalgebra \eqref{sNW} \cite{Concha:2020eam}.

The gauge connection one-form $A$ for the obtained $\overline{\mathfrak{heb}}$ superalgebra is given by \eqref{Asheb}, meanwhile the curvature two-forms read
\begin{align}
    \bar{F}\left(\omega\right)&=F\left(\omega\right)\,,&
\bar{F}^{a}\left(\omega^{b}\right)&=F^{a}\left(\omega^{b}\right)\,,\notag\\
\bar{F}\left(\tau\right)&=F\left(\tau\right)\,,&
\bar{F}^{a}\left(e^{b}\right)&=F^{a}\left(e^{b}\right)+\frac{1}{\ell^{2}}\epsilon^{ac}\tau e_{c}\,,\notag\\
\bar{F}\left(s\right)&=F\left(s\right)\,,&
\bar{F}\left(m\right)&=F\left(m\right)+\frac{1}{2\ell^{2}}\epsilon^{ac}e_{a}e_{c}\,,\notag\\
\bar{F}^{a}\left(k^{b}\right)&=F^{a}\left(k^{b}\right)+\frac{1}{\ell^2}\epsilon^{ac}\tau k_c+\frac{1}{\ell^2}\epsilon^{ac}k e_{c}+\frac{1}{\ell}\bar{\psi}^{+}\gamma^{a}\psi^{-}\,, &
\bar{F}\left(k\right)&=F\left(k\right)+\frac{1}{2\ell}\bar{\psi}^{+}\gamma^{0}\psi^{+}\,,\notag\\
\bar{F}\left(t\right)&=F\left(t\right)+\frac{1}{\ell^2}\epsilon^{ac}k_{a}e_{c}+\frac{1}{2\ell}\bar{\psi}^{-}\gamma^{0}\psi^{-}+\frac{1}{\ell}\bar{\psi}^{+}\gamma^{0}\rho\,, & \bar{F}\left(u_1\right)&=F\left(u_1\right)\,, \notag\\
\bar{F}\left(u_2\right)&=F\left(u_2\right)\,, & \bar{F}\left(b_1\right)&=F\left(b_1\right)\,, \notag\\
\bar{F}\left(b_2\right)&=F\left(b_2\right)\,, \label{eboscurv}
\end{align}
where the $F\left(A\right)$'s are the curvature two-forms \eqref{boscurv} for the $\mathfrak{heb}$ superalgebra. The accommodation of a cosmological constant modifies the $\mathfrak{heb}$ covariant derivatives of the spinor 1-forms as follows
\begin{eqnarray}
\bm{\nabla} \psi^{+}&=&\nabla\psi^{+}+\frac{1}{2\ell}k\gamma_0\psi^{+}+\frac{1}{2\ell}\tau\gamma_0\psi^{+}-\frac{1}{2\ell^{2}}b_1\gamma_0\psi^{+}\,,\notag\\
    \bm{\nabla} \psi^{-}&=&\nabla\psi^{-}+\frac{1}{2\ell}k\gamma_0\psi^{-}+\frac{1}{2\ell}\tau\gamma_0\psi^{-}+\frac{1}{2\ell}k^{a}\gamma_{a}\psi^{+}+\frac{1}{2\ell^{2}}e^{a}\gamma_{a}\psi^{+}+\frac{1}{2\ell^{2}}b_1\gamma_0\psi^{-}\,,\notag\\
    \bm{\nabla} \rho&=&\nabla\rho+\frac{1}{2\ell}k\gamma_{0}\rho+\frac{1}{2\ell}\tau\gamma_0\rho+\frac{1}{2\ell}t\gamma_{0}\psi^{+}+\frac{1}{2\ell^{2}}m\gamma_{0}\psi^{+}+\frac{1}{2\ell}k^{a}\gamma_{a}\psi^{-}\notag\\
    &&+\frac{1}{2\ell^{2}}e^{a}\gamma_{a}\psi^{-}-\frac{1}{2\ell^{2}}b_{1}\gamma_{0}\rho+\frac{1}{2\ell^{2}}b_2\gamma_{0}\psi^{+}\,. \label{efermcurv}
\end{eqnarray}
Here $\nabla\psi^{+}$, $\nabla\psi^{-}$ and $\nabla\rho$ are the corresponding covariant derivatives of the fermionic fields \eqref{fermcurv} for the $\mathfrak{heb}$ superalgebra.

The non-relativistic three-dimensional CS supergravity action based on the $\overline{\mathfrak{heb}}$ superalgebra is derived by considering the gauge connection one-form \eqref{Asheb} and the non-vanishing components of the invariant tensor \eqref{ITSHEB} and \eqref{ITESHEB} into the CS expression \eqref{CSaction},
\begin{eqnarray}
    I_{s\overline{\mathfrak{heb}}}&=&\int \alpha_1\left[ 2e_aR^{a}\left(\omega^{b}\right)-2m R\left(\omega\right)-2\tau R\left(s\right)+k_{a}R^{a}\left(k^{b}\right)-2k \bar{R}\left(t\right)+\frac{1}{\ell^2}e_{a}T^{a}\left(e^{b}\right)\right.\notag\\
    &&\ \left.-\frac{1}{\ell^{4}}\epsilon^{ac}\tau e_{a}e_{c}-\frac{2}{\ell^2}\tau R\left(m\right)+2u_1db_2+2u_2db_1-2\bar{\psi}^{-}\bm{\nabla}\psi^{-}-2\bar{\psi}^{+}\bm{\nabla}\rho-2\bar{\rho}\bm{\nabla}\psi^{+}\right]\notag\\
    &&\ +\alpha_2\left[2k_{a}R^{a}\left(\omega^{b}\right)-2t R\left(\omega\right)-2k R\left(s\right)+\frac{2}{\ell^{2}}k_{a}T^{a}\left(e^{b}\right)-\frac{2}{\ell^{2}}kR\left(m\right)\right.\notag\\
    &&\ \left.-\frac{2}{\ell^{2}}\tau \bar{R}\left(t\right)+\frac{2}{\ell^{2}}b_1db_2-\frac{2}{\ell}\bar{\psi}^{-}\bm{\nabla}\psi^{-}-\frac{2}{\ell}\bar{\psi}^{+}\bm{\nabla}\rho-\frac{2}{\ell}\bar{\rho}\bm{\nabla}\psi^{+}\right]\,,\label{CSseheb}
\end{eqnarray}
where $R^{a}\left(\omega^{b}\right)$, $R\left(\omega\right)$, $R\left(s\right)$ and $R^{a}\left(k^{b}\right)$ are the $\mathfrak{heb}$ curvatures defined in \eqref{acurv} and
\begin{eqnarray}
    T^{a}\left(e^{b}\right)&=&de^{a}+\epsilon^{ac}\omega e_{c}\,,\notag\\
    R\left(m\right)&=&dm+\epsilon^{ac}\omega_{a}e_{c}+\frac{1}{2}\epsilon^{ac}k_{a}k_{c}\,,\notag\\
    \bar{R}\left(t\right)&=&dt+\epsilon^{ac}\omega_{a}k_{c}+\frac{1}{\ell^{2}}\epsilon^{ac}k_{a}e_{c}\,.
\end{eqnarray}
The CS non-relativistic supergravity action \eqref{CSseheb} describes an enlarged Hietarinta extended Bargmann supergravity model which contains three independent sectors. Both the Hietarinta supergravity theory and its enlargement contains the exotic non-relativistic gravity action along $\alpha_0$. In the vanishing cosmological constant limit $\ell\rightarrow\infty$, only the sector proportional to $\alpha_1$ reproduces a supergravity action which is based on the Hietarinta superalgebra. One can see that the inclusion of a cosmological constant term to the Hietarinta supergravity implies the presence of fermionic terms along $\alpha_2$ (see eq. \eqref{CSheb}). The accommodation of a cosmological constant in $\mathfrak{heb}$ framework has implication at the dynamical level where the field equations are given by the vanishing of the curvature two-forms \eqref{eboscurv} and \eqref{efermcurv}. At the bosonic level, the cosmological constant can be seen as a source for a non-vanishing spatial torsion $F^{a}\left(e^{b}\right)=de^{a}+\epsilon^{ac}\omega e_{c}+\epsilon^{ac}\tau\omega_{c}+\epsilon^{ac}kk_{c}$ and for the curvature $R\left(m\right)$. In particular, on-shell we find
\begin{eqnarray}
    F^{a}\left(e^{b}\right)&=&-\frac{1}{\ell^{2}}\epsilon^{ac}\tau e_{c}\,,\notag\\
    R\left(m\right)&=&-\frac{1}{\ell^{2}}\epsilon^{ac}e_{a}e_{c}\,.
\end{eqnarray}
On the other hand, one can notice that the fermionic curvature \eqref{efermcurv} transforms covariantly under the following supersymmetry transformation rules,
\begin{eqnarray}
    \delta \tau&=&\bar{\varepsilon}^{+}\gamma^{0}\psi^{+}\,,\notag\\
    \delta e^{a}&=&\bar{\varepsilon}^{+}\gamma^{a}\psi^{-}+\bar{\varepsilon}^{-}\gamma^{a}\psi^{+}\,,\notag\\
    \delta m&=&\bar{\varepsilon}^{-}\gamma^{0}\psi^{-}+\bar{\varepsilon}^{+}\gamma^{0}\rho+\bar{\eta}\gamma^{0}\psi^{+}\,,\notag\\
    \delta k&=&\frac{1}{\ell}\bar{\varepsilon}^{+}\gamma^{0}\psi^{+}\,,\notag\\
    \delta k^{a}&=&\frac{1}{\ell}\bar{\varepsilon}^{+}\gamma^{a}\psi^{-}+\frac{1}{\ell}\bar{\varepsilon}^{-}\gamma^{a}\psi^{+}\,,\notag\\
    \delta t&=&\frac{1}{\ell}\bar{\varepsilon}^{-}\gamma^{0}\psi^{-}+\frac{1}{\ell}\bar{\varepsilon}^{+}\gamma^{0}\rho+\frac{1}{\ell}\bar{\eta}\gamma^{0}\psi^{+}\,,\notag\\
    \delta b_1&=&\bar{\varepsilon}^{+}\gamma^{0}\psi^{+}\,,\notag\\
    \delta b_2&=&\bar{\varepsilon}^{-}\gamma^{0}\psi^{-}+\bar{\varepsilon}^{+}\gamma^{0}\rho+\bar{\eta}\gamma^{0}\psi^{+}\,,\notag\\
    \delta \psi^{+}&=&D_{\omega}\varepsilon^{+}-\frac{1}{2}u_{1}\gamma_0\varepsilon^{+}+\frac{1}{2\ell}k\gamma_0\varepsilon^{+}+\frac{1}{2\ell}\tau\gamma_0\varepsilon^{+}-\frac{1}{2\ell^{2}}b_1\gamma_0\varepsilon^{+}\,,\notag\\
    \delta\psi^{-}&=&D_{\omega}\varepsilon^{-}+\frac{1}{2}\omega^{a}\gamma_a \varepsilon^{+}+\frac{1}{2}u_1\gamma_0 \varepsilon^{-}+\frac{1}{2\ell}k\gamma_0\varepsilon^{-}+\frac{1}{2\ell}\tau\gamma_0\varepsilon^{-}+\frac{1}{2\ell}k^{a}\gamma_{a}\varepsilon^{+}\notag\\
    &&+\frac{1}{2\ell^{2}}e^{a}\gamma_{a}\varepsilon^{+}+\frac{1}{2\ell^{2}}b_1\gamma_0\varepsilon^{-}\,,\notag\\
    \delta\rho&=&D_{\omega}\eta+\frac{1}{2}\omega^{a}\gamma_a \varepsilon^{-}+\frac{1}{2}s\gamma_0\varepsilon^{+}-\frac{1}{2}u_1\gamma_0 \eta+\frac{1}{2}u_2\gamma_0 \varepsilon^{+}+\frac{1}{2\ell}k\gamma_{0}\eta+\frac{1}{2\ell}\tau\gamma_0\eta+\frac{1}{2\ell}t\gamma_{0}\varepsilon^{+}\notag\\
    &&+\frac{1}{2\ell^{2}}m\gamma_{0}\varepsilon^{+}+\frac{1}{2\ell}k^{a}\gamma_{a}\varepsilon^{-}+\frac{1}{2\ell^{2}}e^{a}\gamma_{a}\varepsilon^{-}-\frac{1}{2\ell^{2}}b_{1}\gamma_{0}\eta+\frac{1}{2\ell^{2}}b_2\gamma_{0}\varepsilon^{+}\,, \label{TR}
\end{eqnarray}
with $D_{\omega}\phi=d\phi+\frac{1}{2}\omega\gamma^{0}\phi$. Here the gauge parameters $\varepsilon^{\pm}$ and $\eta$ are related to the fermionic charges $Q^{\pm}_{\alpha}$ and $R_{\alpha}$, respectively. Let us note that the supersymmetry transformation rules for the Hietarianta extended Bargmann supergravity are derived in the vanishing cosmological constant limit $\ell\rightarrow\infty$.

Thus, the non-relativistic supergravity action \eqref{CSseheb} manifests the existence of an alternative three-dimensional non-relativistic supergravity theory different to the extended Newton-Hooke supergravity and further generalizations presented in \cite{Ozdemir:2019tby,Concha:2020tqx,Concha:2021jos}. The supergravity model derived in this work contrasts with other non-relativistic supergravity theories incorporating a cosmological constant by exhibiting a non-zero spatial super-torsion while maintaining zero spatial curvature $R^{a}\left(\omega^{b}\right)$. Our result could correspond to the torsional subcase of a supersymmetric extension of the non-relativistic Mielke-Baekler (MB) gravity model recently constructed in \cite{Concha:2023ejs}.

\section{Concluding remarks}\label{sec5}

In this work, we have explored the non-relativistic regime of the Hietarinta CS (super)gravity theory in three-dimensional spacetime. Using the contraction and expansion method \cite{Izaurieta:2006zz}, we derived the non-relativistic (super)algebras under the non-degeneracy condition. Our analysis highlights the differences between our Hietarinta model regarding the Maxwellian one \cite{Aviles:2018jzw,Concha:2019mxx}, even though both exhibit isomorphism at the bosonic level. Notably, in the presence of supersymmetry, our theory extends the extended Bargmann supergravity \cite{Bergshoeff:2016lwr} by introducing an additional gauge field that turns on the spatial component of the torsion. These findings present an alternative approach to integrating non-vanishing (super)-torsion in a non-relativistic context, differing from the torsional Newton-Cartan (super)gravity frameworks \cite{Bergshoeff:2015ija,Bergshoeff:2017dqq,VandenBleeken:2017rij}. 

Our results and methodology serve as a starting point for a variety of future studies. An intriguing direction is to investigate the relationship between our work and the recently introduced non-relativistic MB gravity \cite{Concha:2023ejs}. As shown in \cite{Concha:2023ejs}, non-relativistic teleparallel gravity emerges as a specific subcase of the non-relativistic MB gravity where the spatial component of the torsion is non-zero. It would be valuable to explore whether incorporating an additional gauge field into this non-relativistic MB framework could reproduce the Hietarinta extended Bargmann and its cosmological extension in appropriate limits. Furthermore, investigating whether non-vanishing time component of torsion can be derived from a non-zero spatial torsion could offer a novel CS approach to the non-vanishing torsion condition in Newton-Cartan (super)gravity \cite{Bergshoeff:2015ija,Bergshoeff:2017dqq,VandenBleeken:2017rij}. 

Although it is not the purpose of the paper, one might wonder if the Hietarinta approach  to include a non-zero torsion in a non-relativistic regime could be relevant in the effective field theory description of the
quantum Hall effect. It is noteworthy that the non-vanishing torsional condition, first encountered in the context of Lifshitz holography \cite{Christensen:2013lma} and later applied to the Quantum Hall Effect \cite{Geracie:2015dea}, is particularly relevant in CFT settings. Another interesting question is whether the Holographic correspondence can be extended to the Hietarinta non-relativistic realm in which the obtained Hietarinta extended Bargmann (super)gravity in the bulk would be linked to a non-relativistic (super)conformal field theory at the boundary. In the spirit of the AdS/CFT conjecture \cite{Maldacena:1997re}, one could analyze the existence of a connection between the asymptotic symmetry of the Hietarinta CS gravity, described by an extension of the $\mathfrak{bms}_{3}$ algebra \cite{Concha:2023nou}, and a Hietarinta version of the two-dimensional conformal extension of the Galilean algebra ($\mathfrak{gca}_{2}$). This would reveal an holographic duality beyond the known $\mathfrak{bms}_{3}/\mathfrak{gca}_{2}$ correspondence \cite{Bagchi:2010zz}.

Building on the original motivation of Hietarinta \cite{Hietarinta:1975fu}, there are several promising directions for further exploration by incorporating higher-spin content into both relativistic and non-relativistic Hietarinta theories. At the relativistic level, an intriguing avenue would be to study the introduction of spin-3 gauge fields into the Hietarinta gravity, which would generalize the spin-3 extension of the three-dimensional Poincaré gravity \cite{Campoleoni:2010zq}. In the non-relativistic regime, it would be worthwhile to explore how the Hietarinta theory can be generalized to include the spin-3 extended Bargmann gravity model \cite{Bergshoeff:2016soe,Concha:2022muu,Caroca:2022byi}. Additionally, examining the coupling of three-dimensional Hietarinta gravity to a spin-$5/2$ gauge field, in line with the hypergravity approach \cite{Aragone:1983sz,Zinoviev:2014sza,Fuentealba:2015jma,Fuentealba:2015wza,Henneaux:2015ywa,Caroca:2021bjo}, could offer new insights. It would be then particularly interesting to compare the differences between the original Hietarinta-type symmetries \cite{Hietarinta:1975fu} and those emerging from these alternative approaches. 

The non-relativistic expansion involving semigroups presents a promising approach for tackling the intricate task of developing non-relativistic supergravity theory in higher-dimensional spacetime \footnote{A consistent non-relativistic supergravity in D=10 was recently proposed in \cite{Bergshoeff:2021tfn, Bergshoeff:2022iyb} by considering a stringy Newton-Cartan geometry.}. Considering this expansion procedure could be particularly valuable for constructing a supersymmetric extension of the four-dimensional non-relativistic gravity model introduced in \cite{Concha:2022jdc}. However, it is anticipated that the Hietarianta supersymmetry might be broken down to a Nappi-Witten like subalgebra, as we are currently investigating [work in progress].

\section*{Acknowledgment}
The authors would like to thank Javier Matulich for enlightening discussions. P.C. acknowledges financial support from the National Agency for Research and
Development (ANID) through Fondecyt grants No. 1211077 and 11220328. E.R.
acknowledges financial support from ANID through SIA grant No. SA77210097
and Fondecyt grant No. 11220486. P.C. and E.R. would like to thank to the
Direcci\'{o}n de Investigaci\'{o}n and Vicerector\'{\i}a de Investigaci\'{o}%
n of the Universidad Cat\'{o}lica de la Sant\'{\i}sima Concepci\'{o}n,
Chile, for their constant support. S.S. acknowledges financial support from
Universidad de Tarapac\'{a}, Chile.


\bibliographystyle{fullsort}
 
\bibliography{NRHS}

\end{document}